\documentclass[journal=jctcce,manuscript=article]{achemso}

\author{J. Patrick Zobel}
\email{patrick.zobel@univie.ac.at}
\affiliation{Institute of Theoretical Chemistry, Faculty of Chemistry, University of Vienna, W\"ahringerstr.~19, 1090 Vienna, Austria}
\author{Leticia Gonz\'alez}
\email{leticia.gonzalez@univie.ac.at}
\affiliation{Institute of Theoretical Chemistry, Faculty of Chemistry, University of Vienna, W\"ahringerstr.~19, 1090 Vienna, Austria}
\alsoaffiliation{Vienna Research Platform on Accelerating Photoreaction Discovery, University of Vienna, W\"ahringerstr.~19, 1090 Vienna, Austria}

\title{The quest to simulate excited-state dynamics of transition metal complexes}

\usepackage{natbib}
\usepackage[version=3]{mhchem}

\usepackage{xcolor}

\usepackage{graphicx}

\usepackage{natbib}
\usepackage[version=3]{mhchem}
\usepackage{xcolor}
\usepackage{physics}

\renewcommand{\op}[1]{\hat{\mathcal{#1}}}

\begin{document}

\begin{abstract}
This Perspective describes current computational efforts in the field of simulating photodynamics of transition metal complexes.
We present the typical workflows and feature the strengths and limitations of the different contemporary approaches.
From electronic structure methods suitable to describe transition metal complexes to approaches able to simulate their nuclear dynamics under the effect of light, we lay particular attention to build a bridge between theory and experiment by critically discussing the different models commonly adopted in the interpretation of spectroscopic experiments and the simulation of particular observables. 
Thereby, we review all the studies of excited state dynamics on transition metal complexes, both in gas phase and in solution from reduced to full dimensionality.\\
KEYWORDS: \textit{Transition Metal Complexes, Photochemistry, Excited-State Dynamics}
\end{abstract}

\section{Introduction}\label{sec:introduction}
Transition metal complexes offer a rich photochemistry that can be utilized in applications from solar energy conversion to medicine.\cite{Evans2013AP}
This is possible due to the large variety of electronic states of distinct nature that transition metal complexes have to offer. 
For instance, while long-lived states of metal-to-ligand charge transfer  character are key to applications in dye-sensitized solar cells,\cite{ORegan1991NAT,Ponseca2017CR, Wenger2018JACS} short-lived metal-centered states can mediate dissociation processes in biology.\cite{Reinhard2021NAT}
In general, the behavior of transition metal complexes after light irradiation is controlled by the presence or absence of radiationless reaction pathways.
These can either enable efficient transfer between electronic states or facilitate long-lived excited-states that at last may emit. 
Unveiling non-radiative reaction pathways is therefore key to understanding and ultimately tuning the photochemistry of transition metal compounds.

In general, radiationless transitions between electronic states in molecules are categorized into two types: internal conversion --for transitions between electronic states of the same spin multiplicity-- and intersystem crossing, which connects electronic states of different spin multiplicity.
The non-radiative behavior of photoactivated molecules is driven by the motion of the nuclei. 
Upon excitation, the molecules in their equilibrium geometry are lifted from the electronic ground-state potential into an excited-state one.
This brings the molecule to a non-equilibrium situation that induces nuclear motion towards other conformational regions.
On its following journey, the molecule then can pass regions with high probability of non-radiative transitions to other electronic states that ultimately decay back to the electronic ground state.
Alternatively, the molecule can end up in regions with no possibility of (further) non-radiative transfers, from where it can only return to the ground state via luminescence.

Based on the nature of the processes, the study of radiationless pathways experimentally has two prerequisites.
One is identifying a property that changes during the reaction in order to detect which species are present during the reaction.
Since radiationless processes involve a change of the electronic state, these properties may directly or indirectly relate to the electronic potentials --this is the case of following changes in absorption intensities or shifts in vibrational frequencies.
The second is to be able to monitor this property on the same time scale as the reaction occurs.
This time scale is dictated by the nuclear motion of the molecule which takes place typically in the femtosecond regime.\cite{Zewail2000ACIE}
Thus, it is not surprising that electron spectroscopy techniques, such as transient-absorption spectroscopy or time-resolved emission spectroscopy as well as vibrational spectroscopic techniques, such as time-resolved infra-red spectroscopy with sub-picosecond resolution have made incredible progress to access the nature of the ultrafast processes during the non-radiative reactions.\cite{Maiuri2020JACS}

While substantial information can be obtained from spectroscopic experiments, often this might not be sufficient to derive a detailed description of the excited-state dynamics.
Furthermore, with increasing system size, a larger number of vibrational degrees of freedom and greater density of electronic states may participate in the photodynamics, making it more and more difficult to interpret experimental signals.
Most transition metal complexes fall into this category: 
even small coordination complexes already possess dozens of atoms which add up to hundreds of vibrational degrees of freedom and possess many close-lying electronic states due to the only partially filled $d$ shell.
For this reason, experimental studies are almost routinely accompanied by theoretical calculations.\cite{Daniel2015CCR,Daniel2006EIBC,Escudero2019, Penfold2018CR, Daniel2021PCCP}
Most of these efforts involve quantum-chemical calculations of electronic states and relevant points in their potential energy surfaces (PES).
For example, with the help of calculated absorption spectra one can infer the electronic states that may play a role in the photodynamics.
Naturally, it would be better to go beyond this static description and carry out explicit dynamics simulations that directly monitor the time evolution of the electrons and nuclei in the molecule occurring after light irradiation.

One could go as far as to assert that theory not only helps the interpretation of experiments but is a predictive tool on its own.
In practice, however, photodynamics simulations of transition metal complexes are still very much limited due to their computational cost.
This limitation restricts the size of the molecules that can be studied and compromises its accuracy because it necessitates diverse approximations. 
In this predicament one should ponder the benefits of experiment versus theory as follows. 
The experiment can be considered an exact instrument (within experimental resolution) that probes the full system of the molecule interacting with its real environment.
In contrast, theory is an approximate instrument with an inherent methodological error and is often restricted to a truncated system, i.e., reduced molecular model system with or without an environment.
The experiment, however, gives only limited information as one monitors a signal, or --in other words --the global response of the molecule to light. 
In contrast, theory is able to yield every single deactivation pathway. 

Clearly, these contrasting advantages and shortcomings ask for synergy between experiment and theory. 
To efficiently collaborate, practitioners of each side should understand the mindsets and workflows of the other side to adequately assess the information that can be derived from experiment and theory, respectively.
This is particularly important in the developing research field of transition metal photodynamics, where each single discipline is left in the dark when on its own. 

The motivation of this Perspective is illuminating what can be done from the theoretical front (i.e., how much can experiment expect from theory), explain which are the current strategies of simulating the photodynamics of transition metal complexes and its limitations, help in the interpretation of theoretical simulations and pose the challenges still present in the field, 
thereby reviewing the studies done until now. 
If the gap between theory and experiment can be made smaller, more efficient synergies between the two "approaches” can better meet the current and future challenges in transition metal photochemistry.
To this ambitious goal, we present the basic theory and typical ingredients
that enter the simulation of excited-state dynamics simulations, with their strengths and weaknesses, in relation to transition metal complexes.

\section{Theoretical Background}\label{sec:theo}
\subsection{The Time-Dependent Schr\"odinger Equation}\label{sec:sep}

In order to understand the limitations of current theories in the simulation of the photodynamics of transition metal complexes, we deem necessary to introduce some underlying working equations. 
The key equation to solve is the time-dependent Schr\"odinger equation, 
%
\begin{equation}
    i\hbar\frac{\partial\Psi(r,R,t)}{\partial t} = \op{H}^{tot}\Psi(r,R,t)\label{eq:tdse}
\end{equation}
that describes the time evolution of a molecule.
In general, the time-dependent molecular wave function $\Psi(r,R,t)$ depends on the time $t$, the positions $r$ of the electrons, and the positions of the nuclei $R$.
The total Hamiltonian $\op{H}^{tot}$ can be written as 
\begin{equation}
    \op{H}^{tot}= \op{H}^{mol}(r,R) 
    - \hat{\mu}(r,R)\mathbf{\varepsilon}(t) \label{eq:hamil}
\end{equation}
where $\op{H}^{mol}(r,R)$ is the molecular Hamiltonian that contains the kinetic and potential energy terms of the electrons and nuclei in the molecule, as well as a  term $\op{H}^{SOC}(r,R)$ that couples electronic states of different multiplicity via relativistic spin-orbit coupling (SOC) and will be discussed later.
The interaction between the molecule and the light is given in the semiclassical approximation via the dipole operator $\hat{\mu}(r,R)$ and the time-dependent electric field $\varepsilon(t)$.  
Given the complexity of eq.~\ref{eq:tdse}, usually one separates the motion of the nuclei from that of the electrons.
To do this, we can express the molecular wave function 
as a product of electronic wave functions $\Phi^{el}(r;R)$ and nuclear wave functions $\Psi^{nuc}(R,t)$
\begin{equation}
    \Psi(r,R,t)=\sum_{i}\Phi^{el}_i(r;R)\Psi^{nuc}_i(R,t)\label{eq:bornhuang}
\end{equation}
The electronic wave functions $\Phi^{el}(r;R)$ are eigenfunctions of the electronic Hamiltonian $\op{H}^{el}$ that can be obtained through the time-independent electronic Schr\"odinger equation
\begin{equation}
    \op{H}^{el}\Phi^{el}_i(r;R) = E^{el}_i(R)\Phi^{el}_i(r;R)\label{eq:elec_tise}
\end{equation}
The time-independent electronic Schr\"odinger equation can be solved for fixed selections of nuclear coordinates $R$ representing a specific molecular geometry.
Thus, the electronic wave functions (and so its properties) depend only parametrically on the nuclear coordinates $R$, and their eigenvalues $E^{el}(R)$ yield the potential energy for this specific geometry.

As a result, the motion of the nuclear wavefunctions $\Psi^{nuc}_i$ in the electronic potential $E^{el}_i$ is described by
\begin{equation}
    \op{H}_{i}^{mol}=E^{el}_i (R) + \op{T}_{i}^{nuc}
    \label{eq:hii}
\end{equation}
while the electronic potentials are coupled through off-diagonal terms that include
\begin{equation}
    \op{H}_{ij}^{mol}=\op{T}_{ij}^{NAC} + \op{H}^{SOC}(R)
    - \hat{\mu}_{ij}(R)\mathbf{\varepsilon}(t)
    \label{eq:hij}
\end{equation}
The term $\op{T}_{ij}^{NAC}$ are the so-called non-adiabatic couplings (NACs), defined as  
\begin{align}
  \op{T}^{NAC}_{ij} =& -\sum_{Nuclei\;A}\frac{1}{2M_A}\Big[ 
   \left\langle\Phi_j^{el}(r;R)|\nabla^2_{R_A}|\Phi_i^{el}(r;R)\right\rangle\nonumber\\ &+ \left\langle\Phi_j^{el}(r;R)|\nabla_{R_A}|\Phi_i^{el}(r;R)\right\rangle\nabla_{R_A} \Big] \label{eq:nac}
\end{align}
that furthermore couple the motion of the nuclei and electrons.
$\op{\mu}_{ij}$ is the transition-dipole moment between electronic states $\Phi^{el}_i$ and $\Phi^{el}_j$, and $\nabla_{R_A}$ denotes the gradient with respect to the nuclear coordinates $R_A$.

\subsection{The Non-Adiabatic Couplings}\label{sec:theo_nac}
In electronic structure theory, the molecular Schr\"odinger equation is solved (yet approximately) only by neglecting the NACs of eq.~\ref{eq:nac}. 
In this Born-Oppenheimer (or adiabatic) approximation, the motion of the nuclei and electrons is completely decoupled and the nuclear wave functions are then constrained to a single electronic potential.
Naturally, this approximation is only useful when describing processes that take place in a single electronic state.
This can be the case, when the electronic state in question is well separated in energy from all other electronic states, e.g., as is usually found for reactions that occur in the electronic ground state.
However, if we are interested in reactions including several electronic states, it is mandatory to include NACs to allow the nuclear wave function to transfer between the different electronic states (internal conversion).
In eq.~\ref{eq:nac}, the second term in the sum can be written as
\begin{equation}
    \mel{\Phi_j^{el}(r;R)}{\nabla_{R_A}}{\Phi_i^{el}(r;R)}= \frac{1}{E_j-E_i}
    \mel{\Phi_j^{el}(r;R)}{\nabla_{R_A}\op{H}^{el}}{\Phi_i^{el}(r;R)}\label{eq:hellmann}
\end{equation}
This shows that the NACs become large when the corresponding electronic states are close in energy ($E_j- E_i\approx0$).
Since it is the NACs that enable the transfer between different electronic states, state transfer is most efficient when the corresponding states are close in energy.

\subsection{The Spin-Orbit Couplings}\label{sec:theo_soc}
The SOC couples electronic states of different spin multiplicities and, thus, allows both radiative transitions (phosphorescence\cite{Baryshnikov2017CR}) and non-radiative transitions (intersystem crossing\cite{Penfold2018CR, Marian2021ARPC}) between them.
SOC is a relativistic effect, as it occurs naturally in a formulation of quantum mechanics that includes the principles of the theory of special relativity.
Phenomenologically, SOC is explained as the interaction of the magnetic moment of the spin angular momentum with the magnetic field that is induced by the the electron orbiting around the nuclei as well as in the field of the other electrons.\cite{Marian2001, Marian2012WIRES}
In non-relativistic quantum theory based on the Schr\"odinger equation, SOC, has to be introduced ad hoc.
Using the  Breit-Pauli operator, $\op{H}^{SOC}$ 
can be expressed as
\begin{align}
  \op{H}^{SOC} =& \frac{1}{2m_ec^2}\sum_{Electrons\;i}\left[
   \sum_{Nuclei\;A}Z_A\left(\frac{\hat{r}_{iA}\times\hat{p}_i}{\hat{r}_{iA}^3} \right)\cdot\hat{s}_i \right.\nonumber\\
   &+\left. \sum_{Electrons\;j\neq i}\left(
\frac{\hat{r}_{ij}\times\hat{p}_i}{\hat{r}^3_{ij}}
    \right)\cdot\left(\hat{s}_i+2\hat{s}_j \right)
  \right]\label{eq:bp}
\end{align}
%

The first sum in eq.~\ref{eq:bp} describes the interaction of each electron's orbital angular momentum $\hat{r}_{iA}\times \hat{p}_i$ (orbiting around the nucleus $A$) with its spin $\hat{s}_i$.
This interaction depends on the charge of the nucleus $Z_A$.
For valence electrons in many-electron systems, SOCs typically scale as $Z_A^2$,\cite{Pyykko2012ARPC}, which manifests itself strongly in heavy atoms.
In transition metal complexes, SOCs are often strong enough to allow ultrafast intersystem crossing,\cite{Chergui2012DT} and it is thus necessary to include them during nonadiabatic dynamics simulations.
The strength of the SOC is not only dependent on the presence of heavy atoms in the molecule, but is also depends on the character of the electronic states that are coupled. 
This is rationalized by the generalized El-Sayed rules\cite{Marian2021ARPC}, that state that in order to provide large SOCs for efficient intersystem crossing, (i.) the coupled electronic states should differ only by a single excitation, which (ii.) involves a change in orbital type, and (iii.) the orbitals should be localized at the same site in the molecule.

\begin{figure*}[ht]
    \centering
    \includegraphics[width=\textwidth]{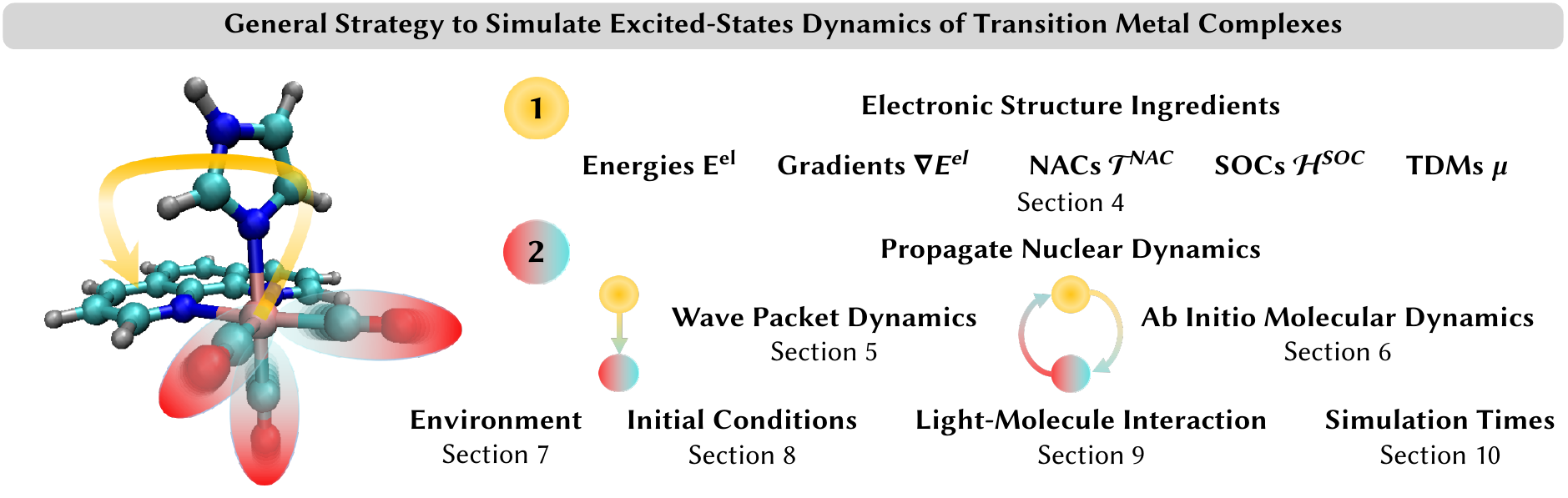}
    \caption{General strategy to simulate excited-state dynamics of transition metal complexes, as discussed in this Perspective. Step 1 refers to the ingredients that must be calculated from electronic structure theory. Step 2 embraces the two strategies employed for photodynamics of transition metal complexes so far: wave packet dynamics and ab initio molecular dynamics.  
    $E^{el}$: electronic energies; NACs: non-adiabatic couplings (eq.~\ref{eq:nac}); SOCs: spin-orbit couplings (eq.~\ref{eq:bp}); TDMs: transition-dipole moments.}
    \label{fig:strategy}
\end{figure*}

\section{The General Strategy to Simulate Excited-State Dynamics}\label{sec:elec_practical}
From the previous section, 
it is clear that in order to follow the time-evolution of a molecule, we first need to collect several electronic ingredients: the electronic energies $E^{el}_i(R)$ (eq.~\ref{eq:elec_tise}) and implicitly their gradients, the NACs (eq.~\ref{eq:nac}), the SOCs (eq.~\ref{eq:bp}) and eventually also the transition dipole moment $\mu_{ij}(R)$ if the light-matter interaction is explicitly simulated (eq.~\ref{eq:hij}).
In the second step, the motion of the molecule can be simulated by propagating the nuclear motion in the different (coupled) electronic states.
This strategy, which guides the structure of the remainder of this Perspective, is collected in Figure~\ref{fig:strategy}.

Both the electronic and nuclear steps face practical challenges.
On the one side, enough accuracy in the electronic structure part is needed.
On the other side, the scaling of the computational costs inherent to the size of the molecule need to be controlled.  
In the exact limit, this cost scales exponentially with the molecular size for solving both the electronic and the nuclear problem.
To alleviate the steep scaling in the electronic part, a large number of electronic structure methods have been developed whose scaling follows different power laws.
Thus, when choosing one electronic structure method for dynamics, the selection is strongly motivated by maximizing accuracy for a given computational effort.
This situation is different in the nuclear part, where 
selecting a method is more a fundamental choice of which processes should be qualitatively well described, with less focus on quantitative accuracy.
A closer insight of the available possibilities to solve the electronic and nuclear problems is given in Section ~\ref{sec:elec} and Sections~\ref{sec:wp}-\ref{sec:aimd}, respectively.

\section{Electronic Structure Methods for Dynamics of Transition Metal Complexes}\label{sec:elec}
\subsection{Practical Considerations}
There exist a large number of both commercial
and (academically) freely distributed
quantum chemistry computer program packages that offer a wild range of electronic structure methods to calculated the electronic ingredients.
In practice, however, the selection of a particular method to underlay non-adiabatic nuclear dynamics is limited to those implementations that are able to provide the energies, SOCs and possibly gradients and NACs, depending on the approach chosen for dynamics.

When gradients are needed, it is desirable to have implementations that allow for analytical rather numerical gradients.\cite{Pulay2013WIRES}
Analytical implementations obtain the derivatives along all nuclear coordinates in a single calculation instead of requiring the "manual" displacement along each nuclear coordinate that makes numerical implementations computationally more expensive; however, 
analytical gradients are not available for many methods. 
When NACs are not available, it is possible to approximate the state-to-state transition probabilities by computing the overlap of the wave functions.\cite{HammesSchiffer1994JCP, Plasser2016JCTC}
The SOC matrix elements can also be calculated a posteriori for some quantum chemical methods.\cite{Gao2017JCTC} 

The choice of a quantum chemical method can be further limited, if solvent is included implicitly in the dynamics simulation, as all required properties need to be available including the solvent or if schemes containing the solvent explicetly are not implemented with the electronic structure method of choice.
Fortunately, steady effort in the development of quantum chemistry program packages is continuously adding new implementations that can be employed for nonadiabatic dynamics simulations.

Depending on the number of nuclear degrees of freedom considered, the number of individual electronic structure calculations that need to be performed for the excited-state dynamics simulations can easily reach $\sim10^5$.
Compared to static explorations of PES with typically only few dozens of calculations, this is an enormous increase in computational cost.
Thus, it is often necessary to balance the trade-off between computational cost and accuracy differently between static and dynamic studies of excited-state processes.
We caution, however, that this does not imply that anything goes!
A dynamics study with wrong electronic ingredients 
will only produce a collection of meaningless numbers, irregardless of how much computational effort has been invested in the dynamics simulation.
Instead, the selection of the electronic structure method has to be guided by carefully testing their performance against experimental reference data or other reliable electronic structure methods.\cite{Mata2017ACIE}

Therefore, it gets clear that besides the question of availability and efficiency, the choice of an electronic structure method should be tailored to describe each transition metal complex and the reaction of interest. 
Since we are interested in electronically excited states, the first experimental reference of choice is often the static absorption spectrum.\cite{Loos2019CPC, Bai2020JMM}
A calculated spectrum with the chosen method should qualitatively match the experimental one in terms of number of observed absorption bands, their relative intensity, and preferably also band shoulders.
For the relative position of the calculated and measured absorption bands, one should strive for energetic differences of 0.1-0.5~eV as this is a reasonable accuracy that can be achieved for excitation energies of medium-sized and larger molecules. 
Smaller average error may be obtained only with very accurate methods; however, such methods are most likely computationally unfeasible except for few-atomic molecules.\cite{Loos2020JPCL}

Sometimes, larger errors in the total excitation energies can be acceptable in excited-state dynamics studies, when the error in the relative excitation energies is small and the same along the PESs of interest.
For example, dynamics excluding relaxation to the ground state might be well described despite larger errors in the excitation energies, if the error is systematic for all excited states, i.e., when all excitation energies are under- or overestimated by a similar amount and can be accordingly scaled.\cite{Frutos2007PNAS, Kinzel2012PCCP}

An additional cross-check can be achieved by computing time-resolved spectra that can be directly compared with experimental counterparts, coming e.g., from time-resolved absorption spectroscopy\cite{Polli2010NAT, Timmers2019NATCOM}, time-resolved emission spectroscopy\cite{Cusati2011JACS, Mai2019CS}, time-resolved photoelectron spectroscopy\cite{Hudock2007JPCA,Mitric2011JPCA, Brogaard2011JPCA, Mai2018MOL}, or time-resolved X-ray scattering.\cite{Kirrander2016JCTC, Papai2019JCP}
Such calculations can a posteriori validate the selection of the electronic structure methods. 

Particularly if no experimental references are available, it is useful to compare the calculated energies and geometries of selected points in the PES --such as ground- and excited-state minima, conical intersections, or interstate crossings-- against other (higher-level) electronic structure methods.
Available benchmark studies can also help in this endeavour.
However, while the number of benchmark studies that systematically investigate excitation energies \cite{Schreiber2008JCP, SilvaJunior2008JCP, SilvaJunior2010JCP, SilvaJunior2010MP, Loos2018JCTC, Loos2019JCTC, Loos2020JCTC, Loos2020JCTCb} or oscillator strengths and excited-state dipole moments\cite{Sarkar2021JCTC} of organic molecules continuously grows, corresponding studies for transition metal complexes are much more scarce and limited in size, e.g., Refs.~\citenum{Rosa2004,Niehaus2015RSCA,Almeida2016,Atkins2017JCTC}.
The absence of such data is not surprising as benchmarking transition metal complexes requires considerable effort due to the larger size and the greater variety of characters of electronic states in transition metal complexes compared to organic molecules.
For some families of wave-function based electronic structure methods, there exists a hierarchy in terms of accuracy,\cite{Mata2017ACIE} which can be consulted in the absence of other reference data to decide on the reliability of the computed PES.

Finally, we note that even if the ground state might not play an apparent role in the excited dynamics simulations, the method of choice should be capable to describe it. 
This is because electronic excited states are described either by a configuration interaction or (coupled-) cluster expansion or from perturbation theory/response theory applied to the electronic ground state.
Thus, at least a qualitatively correct ground-state wave function is mandatory to obtain a correct description of the excited states.
This requirement should be extended to all geometries that may be visited during the dynamics.
Additionally, the quality of the electronic ground state also influences the vertical excitations energies, which in turn can affect the  excitation wavelengths and even the course of the dynamics. 

In general, electronic structure methods can be grouped into two kinds of approaches, multi-configurational (sometimes also grouped together with multi-reference) methods and single-reference methods.\cite{Mai2020ACIE,Gonzalez2021QCES}
These approaches will be discussed in the following Sections~\ref{sec:elec_mr}--\ref{sec:elec_sr} with a focus on transition metal complexes.
Semi-empirical methods, which simplify the solution of eq.~\ref{eq:elec_tise} by replacing expensive integrals with experimentally parameterized corrections,~\cite{Akimov2015CR} deserve an extra category (Section~\ref{sec:elec_semi}), since they can be formulated in SR or MR variants.

\subsection{Multi-Configurational and Multi-Reference Electronic Structure Methods}\label{sec:elec_mr}
\subsubsection{Applicability}
\begin{figure*}
    \centering
    \includegraphics[width=\textwidth]{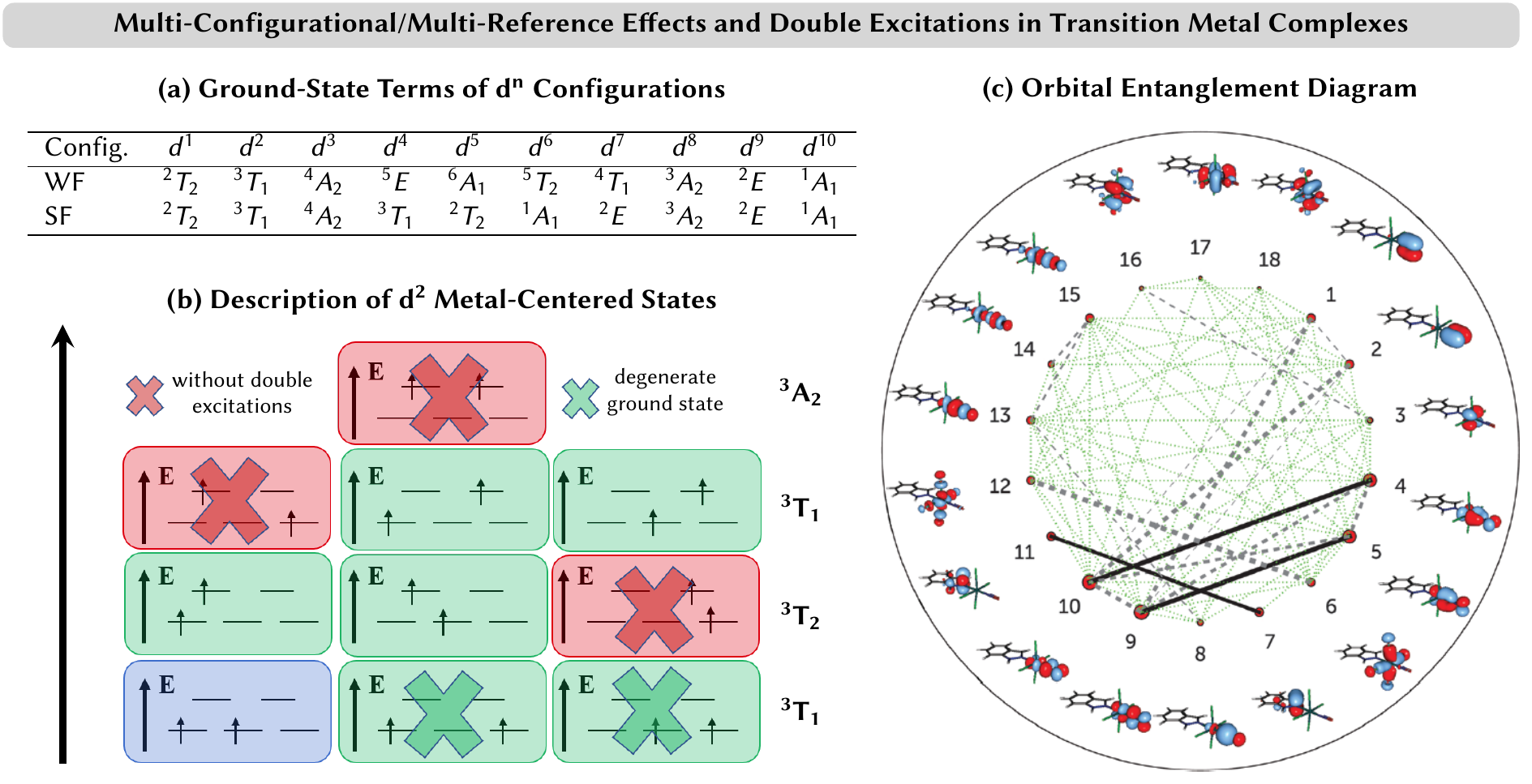}
    \caption{(a) Ground-state terms of $d^n$ configurations for weak (WF) and strong (SF) ligand fields in octahedral complexes. Only terms of $A$ symmetry are spatially non-degenerate and can be fully accounted for by single-reference methods.
    (b) Description of the triplet metal-centered electronic states of an octahedral $d_2$ complex. A single-reference approach selects one component (blue) from the $^3T_1$ ground-state term, while the other two can only be described as (singly) excited states. Based on this selection, some excited-state components are only accessible via double excitations. To describe all states equally well, a multi-configurational/multi-reference method including higher order excitations is needed. Adapted from Ref.~\citenum{Zobel2021CS}. Copyright by the Americal Chemical Society 2021.
    (c) Orbital entanglement diagram of a DMRG calculation using an active space of 18 orbitals for the \textit{trans}-\ce{[RuCl_4(NO)(1H-indazole)]}  complex.  
    The size of the circles denotes the single orbital entropy as a measurement of the involvement of the orbital in open-shell electronic configurations. The thickness of the lines denotes the mutual information as a measurement of the interaction of the orbitals. This exemplifies the importance of all orbitals in the correct description of the excited states of the molecule, highlighting the need for large active spaces in the study of transition metal complexes.
    Image adapted with from Ref.~\citenum{Freitag2015PCCP} under Creative Commons Attribution 3.0 Inportend Licence. Published by the PCCP Owner Societies (2015).
    } 
    \label{fig:multi}
\end{figure*}

In contrast to single-reference (SR) methods,\cite{Dreuw2005CR} which describe the ground state by a single configuration expressed by a Slater determinant, 
multi-configurational (MC) methods\cite{LiManni2021MCQC} consider more than one configuration/Slater determinant and multi-reference (MR) methods\cite{Lischka2018CR, Park2020CR} use such linear combination of multiple configurations/Slater determinants for the ground state to generate further excitations. 
By construction, MR/MC approaches offer a more flexible description than the SR ansatz and they are also capable to describe open-shell electronic ground states, which can frequently be encountered in transition metal complexes due to the partially filled $d$-shell of the metal atom, as well as capable to describe ligand dissociation. 
For example, among octahedral complexes, most $d^n$ configurations (see Figure~\ref{fig:multi}a) feature spatially degenerate electronic ground-state terms that can only be described correctly by using multiple Slater determinants.

\subsubsection{Complete Active Space Self-Consistent Field and Beyond
}
The most popular MC approach for the study of transition metal complexes is the family of complete-active-space self-consistent field (CASSCF)-based methods.\cite{Roos2008MQCGES, Olsen2011IJQC, Roos2016MQC}
The CASSCF method itself\cite{Roos1980CP} is capable of achieving a qualitative correct description where a SR method fails; 
however, CASSCF excitation energies can often bear sizable errors.\cite{Gonzalez2012CPC}
Therefore, it is often used as the starting point of a second-order perturbation theory treatment, e.g., in the CASPT2\cite{Andersson1990JPC} or NEVPT2\cite{Angeli2001JCP} approaches, or in a subsequent configuration interaction (CI) expansion, as in the multi-reference configuration interaction (MRCI)\cite{Szalay2012CR} ansatz.
CASSCF-based methods rely on the selection of an active space, which comprises a subset of occupied and unoccupied orbitals of the molecule. 
For this subset, the best possible wave function is then calculated in a full CI treatment with simultaneous re-optimization of the orbitals. 
Subsequent perturbation or CI expansions of a CASSCF wavefunction add energy corrections to the CASSCF states. 

CASSCF-based calculations scale factorially with the size of the active space.
This scaling imposes a limit to the maximum size of the active space that can be treated nowadays, perhaps of ca.~20 orbitals including 20 electrons, as in the Cr$_3$ example of Ref.\citenum{Vogiatzis2017JCP} using $4\cdot10^9$ Slater Determinants in a single electronic structure calculation.
Therefore, it is important to carefully restrict the active space to the most important orbitals expected to contribute to the problem of interest~\cite{Veryazov2011IJQC, Roos2008MQCGES} such as the orbitals that characterize the ground and excited states in the nonadiabatic dynamics simulations.
For transition metal complexes, this can include the metal $d$ orbitals, metal-ligand-$\sigma$ and $\sigma^\ast$ orbitals, as well as $\pi$, $\pi^\ast$, and non-bonding $n$ ligand orbitals that typically characterize the low-lying excited states.
In addition, for complexes featuring $3d$ metals, it is advised to include a second set of $3d'$ orbitals (a double $d$ shell) when studying processes where the occupation of the $d$ shell changes, such as metal-to-ligand excited states.\cite{Roos2008MQCGES}
One realises that for transition metal complexes, attempting to include all formally important orbitals in the active space quickly exceeds the computational limits of the standard CASSCF ansatz. 
Extended approaches to treat larger active spaces have been developed as the restricted (RASSCF)\cite{Olsen1988JCP} and generalized (GASSCF)\cite{Ma2011JCP} variants of CASSCF or the density matrix renormalization group (DMRG)\cite{Marti2011PCCP, Sharma2012JCP} approach.
The RASSCF/GASSCF methods allow the usage of larger active spaces of ca.~30 orbitals\cite{Ma2011JCP} by defining sub-spaces in the active space with limited interaction between the orbitals in the different sub-spaces. 
The active space limit can be pushed further up to the order of 100 orbitals in DMRG,\cite{Hachmann2006JCP} which uses tensor decomposition methods to approximate the CASSCF wave function.
While for typical DMRG calculations, active space sizes of ca.~50 orbitals are manageable\cite{OlivaresAmaya2015JCP}, when applying a subsequent second-order perturbational treatment to the DMRG wave function as in CASPT2 or NEVPT2 approaches to obtain more accurate electronic energies, the active space size that can be handled computationally decreases again to ca.~30 orbitals\cite{Freitag2021DMRG} --for a single point energy calculation.
Figure~\ref{fig:multi}c shows an example of a so-called entanglement diagram that can be obtained from a DMRG calculation to estimate the importance of different orbitals in a active space.
Other methods to include more than one configuration  can be found in Ref.~\citenum{Gonzalez2021QCES} but, as RASSCF/GASSCF and DMRG, they have not been used for dynamics of transition metal complexes yet.

\subsubsection{Application to Excited-State Dynamics of Transition Metal Complexes}
MC and MR methods have been used in a number of nonadiabatic dynamics studies of dissociation processes of hydrido carbonyl complexes\cite{Guillaumont1999JACS, Heitz1997CCR, Heitz1997JCP, Heitz1997JACS, Daniel1996IJQC, Daniel1994IJQC, Daniel1994JPC, Daniel1993JPC, Ambrosek2007JPCA, Heitz2000JOC} and related compounds.\cite{BruandCote2002CEJ, Paterson2002JPCA,Daniel2003SCI, Worth2006MP, Daniel1994IJQC, Guillaumont1998CCR, Finger1996JPC, Full2003PCCP, Heitz2000JOC, Daniel2001CP, Ambrosek2007JCP, Full2006CP}
These studies are based on wave-packet dynamics simulations (see Section~\ref{sec:wp}) on PES obtained mostly by CASSCF/MRCI calculations with active spaces including up to 14 orbitals.\cite{Daniel2001CP, Ambrosek2007JPCA, Full2003PCCP} 
However, in these cases the PES were restricted to one or two dissociative coordinates. 
Similarly, a wave-packet dynamics study of small rhenium(VII)\cite{Costa2008NJC} and chromium(III)\cite{Ando2012CPL} complexes employed CASSCF by itself and CASSCF combined with quasi-degenerate perturbation theory, respectively, on one-dimensional PES.
In comparison, the number of studies using MR or MC methods in ab initio molecular dynamics simulations (see Section~\ref{sec:aimd}) is much smaller. 
We could find only one study: an ab initio multiple spawning (see Section~\ref{sec:aims}) study of a small iron(II) complex, which included the ground state $S_0$ and the first-excited singlet state $S_1$ computed with CASSCF and an active space including 11 orbitals.\cite{Bera2017JCP}
This scarcity is in strong contrast to ab initio molecular dynamics simulations of organic molecules, where MC/MR methods such as CASPT2 or MRCI are used very often  \cite{Manathunga2016JCTC, Park2017JCTC, Mai2016JPCL, Liu2016JPCB, Mai2016NC} with active spaces including up to 11 orbitals.\cite{Park2017JCTC}

As it will be introduced later (Section ~\ref{sec:wp_lvc}), it is also possible to use parameterized PES where to run nonadiabatic simulations at a much lower cost.
This strategy has been recently employed in a wave-packet dynamics study of a heme-CO complex\cite{Falahati2018NATC} carried out on 15-dimensional CASSCF/CASPT2
PES, as well as in an ab initio molecular dynamics study of a vanadium(III) complex on its full 123-dimensional CASSCF PES.\cite{Zobel2021CS} 
These studies used active spaces including 9 and 13 orbitals, respectively.

\subsection{Single-Reference Electronic Structure Methods}\label{sec:elec_sr}
\subsubsection{Applicability}
When all molecular geometries visited during the dynamics possess an electronic ground state that can be described by a single configuration, then SR electronic structure methods can be employed to calculate the electronic potentials, gradients, and couplings.
This condition might be fulfilled if the molecule possesses a closed-shell electronic ground state, and neither dissociates nor undergoes internal conversion from the first excited singlet state $S_1$ to the ground state $S_0$ during the dynamics simulation. 
Note that the opposite is not true, i.e., a SR method does not necessarily describe a close-shell ground-state molecule well.

Compared to MC/MR, SR methods feature a considerably lower scaling in the computational cost with regard to the system size\cite{Loos2020JPCL}, making them more amenable to transition metal complexes. 
Applications revolve around (closed-shell) low-spin $d^6$ complexes\cite{Fumanal2017JCTC, Fumanal2018JPCL, Fumanal2018PCCP, Eng2015ACR, Gourlaouen2015JCTC, Harabuchi2016JCTC, Papai2016JPCC, Papai2016JPCL, Papai2018JCTC, Papai2019JCP, Papai2019JPCC, Mai2020TCA, Zobel2020IC, Heindl2021IC, Fang2019CTC, Mai2019CS, Atkins2017JPCL, Tavernelli2011CP, Freitag2014IC, Talotta2020CEJ, Liu2018JPCA, Fumanal2021JCP} and $d^{10}$ systems\cite{Capano2014JPCA, Eng2020PCCP, Capano2017PCCP, Giret2021JMCC, Liu2018JCPb} with one single exception.\cite{Dorn2020JACS}
As a further advantage, SR methods can be considered more "black-box" than MC/MR methods as they typically depend on fewer (and less critical) parameters, which makes them more user friendly. 

\subsubsection{Time-Dependent Density Functional Theory}
The most popular and computationally lowest-scaling SR method is time-dependent density functional theory (TDDFT).\cite{Ferre2016DFTES} 
It is based on the standard (time-independent) DFT, in which the ground-state energy of the molecule is expressed as a functional of the electron density.
The energy can be calculated in the Kohn-Sham formalism,\cite{Marques2004ARPC} that assumes the electron density of the real system to be identical to that of a fictitious system of non-interacting electrons.
The electrons in the non-interacting system can conveniently be described exactly by a set of orbitals to calculate the electron density.
This gives the ground-state wave function the form of a single configuration.
The differences between the system of interacting and non-interacting electrons are combined in a so-called exchange-correlation (XC) functional, that also depends on the density.
The drawback of the Kohn-Sham formalism, however, is that the exact form of the XC functional is unknown, which has lead to numerous efforts to find approximated forms of XC functionals.\cite{Mardirossian2017MP}
For excited-state calculations, TDDFT can be cast in a linear-response (LR) formulation\cite{Casida1995} that allows the calculation of excitation energies from matrix eigenvalue problems without the need to propagate the time-dependent density explicitly, and it yields state vectors that are analogous to the CI expansions from wavefunction methods.
TDDFT is thereby typically employed in the adiabatic approximation,\cite{Marques2004ARPC} that uses the XC functionals from ground-state DFT.

In practical applications, TDDFT can suffer from a number of problems,\cite{Casida2012ARPC, Maitra2016JCP} the most notorious being the dependence of the calculated energy and properties on the choice of XC functional.
For excited states of transition metal complexes, hybrid XC functionals, such as B3LYP or PBE0, seems to be preferable compared to generalized-gradient functionals, as discussed, e.g., in Refs.~\citenum{Vlcek2007CCR, Latouche2015JCTC}.
This preference is also mirrored in excited state dynamics studies, that mostly rely on hybrid functionals\cite{Fumanal2017JCTC, Fumanal2018JPCL, Fumanal2018PCCP, Eng2015ACR, Gourlaouen2015JCTC, Harabuchi2016JCTC, Papai2016JPCC, Papai2016JPCL, Papai2018JCTC, Papai2019JCP, Papai2019JPCC, Mai2020TCA, Zobel2020IC, Heindl2021IC, Fang2019CTC, Mai2019CS, Eng2020PCCP, Capano2017PCCP, Giret2021JMCC, Fumanal2021JCP, Liu2018JCPb} with some exceptions.\cite{Atkins2017JPCL, Tavernelli2011CP, Freitag2014IC, Capano2017PCCP, Talotta2020CEJ}
In addition to the general functional dependence, TDDFT is well known to suffer from its inability to describe CT excitations with standard XC functionals.\cite{Dreuw2003JCP}
This problem is alleviated using long-range corrected XC functionals.\cite{Yanai2004CPL, Chai2008JCP}
For transition metal complexes, the CT problem is more prominent for interligand CT excitations, while CT transitions between metal and ligands are usually described well.\cite{Daniel2015CCR}
When a large number of excited states is calculated, TDDFT can fail to describe high-energy states with energies between the ionization potential and the negative energy of the highest-occupied molecular orbital correctly.
This problem, however, can be corrected using special asymptotically-corrected XC functionals.\cite{Casida1998JCP}
Finally, TDDFT describes excited states only using single excitations.
This not only disregards double and higher-order excitations (see an example in Figure~\ref{fig:multi}b), but can also fail to describe single excitations when they involve a spin-flip in open-shell molecules.\cite{Casida2005JCP}
Some states of double excitation character are, in turn, accessible using a spin-flip TDDFT ansatz.\cite{Shao2003JCP}
As is summarized in Ref.~\citenum{Casida2012ARPC}: TDDFT works best for "low-energy one-electron excitations involving little or no charge transfer and that are not too delocalized".
In all other cases, care should be exercised.

\subsubsection{Coupled Cluster and Related Methods}
Besides TDDFT, there exist other SR wave-function methods, most notable coupled cluster (CC) \cite{Sneskov2012WIRES} as well as the algebraic diagrammatic construction (ADC) scheme of the polarization propagator.\cite{Dreuw2015WIRES}
Both methods have a clear hierarchy that allows to  improve the accuracy of the results systematically with a concomitant increase of computational cost. 
The most economic variants, approximate second-order CC (CC2)\cite{Christiansen1995CPL} and second-order ADC [ADC(2)]\cite{Schirmer1981PRA}, scale slightly larger than TDDFT with the system size ($N^5$ vs.~$N^4$) and can deliver similar accurate excitation energies\cite{Harbach2014JCP, Jacquemin2015JCTC} as well as excited-state geometries\cite{Hattig2005AQC} at least for organic molecules, provided the excited states are dominated by single excitations.\cite{Dreuw2015WIRES}
In contrast to TDDFT, both CC2 and ADC(2) variants include double excitations, however, only at a zeroth-order level.
This contributes some admixture of single and double excited states, but it is insufficient to describe states with large double-excitation character adequately.
Probably for this reason, CC2 or ADC(2) studies on transition metal complexes are rare\cite{Escudero2015PCCP} and in general little reliable.

\subsection{Semi-Empirical Methods}\label{sec:elec_semi}

Semi-empirical methods\cite{Akimov2015CR} usually neglect or parametrize molecular integrals occurring in the calculation of the electronic Schr\"odinger equation (eq.~\ref{eq:elec_tise}).
This way they can deal with large molecules that might be prohibitive with ab initio wavefunction theory or even DFT. 
Semi-empirical methods can be based either on molecular orbital (MO) theory\cite{Bredow2005TCA} or on DFT.\cite{Elstner2014PTRSA}
Further, they can be combined with CI schemes\cite{Liu2018JCP} or MRCI ones~\cite{Tuna2016JCTC}  to be used in ab initio molecular dynamics simulations.

However, while semi-empirical MO methods can reproduce excitation energies of small and medium-sized organic molecules with deviations of 0.4-0.5~eV 
\cite{SilvaJunior2010JCTC}, benchmarks for excited states of transition metal complexes are missing.
Even for the prediction of ground-state energetics of transition metal complexes, semiempirical MO methods can struggle when tested against DFT results.\cite{Minenkov2018JCTC} 
As a matter of fact, we are not aware of any excited state dynamic study of transition metal complexes using semi-empirical MO methods.

A representative of DFT-based semiempirical methods is density-functional tight-binding (DFTB).
DFTB relies on a truncated Taylor expansion  of the DFT energy with respect to the fluctuation of the electron density around a reference density, which is typically given by a superposition of atomic densities.\cite{Akimov2015CR}
In analogy to the parent DFT approach, DFTB can be used in a linear-response time-dependent formulation to calculate excited states and has been used in this manner in excited state simulations.\cite{Stojanovic2017JCTC}
Excitation energies of small and medium-sized organic molecules obtained with linear-response time-dependent DFTB can reproduce TDDFT results using the PBE functional with deviations of ca.~0.2~eV.\cite{Rueger2016JCP}
Furthermore, DFTB approaches can reproduce DFT geometries of transition metal complexes reasonably.\cite{Zheng2007JCTC}
Even if benchmark studies of DFTB excitation energies are missing for transition metal complexes, excited state dynamics studies of transition metal complex should emerge in the near future.

\section{Nuclear Quantum Dynamics Methods}\label{sec:wp}
\subsection{Wave Packet Propagation on a Grid}\label{sec:wp_standard}

Having selected a suitable method to compute the electronic-structure ingredients, we can now tackle the simulation of the nuclear motion given by the time-dependent nuclear Schr\"odinger equation,
\begin{equation}
    i\hbar\frac{\partial\Psi^{nuc}(R,t)}{\partial t} = \op{H}^{tot}\Psi^{nuc}(R,t)\label{eq:tdse_nuc2}
\end{equation}
where the nuclear and electron degrees of freedom are separated, as explained in Section~\ref{sec:sep}.
Integrating this equation, one can obtain\cite{Reiter2021} the nuclear wave function
\begin{equation}
    \Psi^{nuc}(R,t)=\op{U}(t,t_0)\Psi^{nuc}(R,t_0)=\exp^{-i\op{H}^{tot}(t-t_0)/\hbar}\Psi^{nuc}(R,t_0)\label{eq:propagator}
\end{equation}
where $\op{U}(t,t_0)$ is a propagator that evolves the wave function from the initial time $t_0$ to the final time $t$.
Often, the so-called split-operator method\cite{Fleck1976AP} is used to evaluate numerically this propagator.

The general solution for eq.~\ref{eq:tdse_nuc2} is a nuclear wave function represented as a linear combination of specific time-independent basis functions, also known as a wave packet.

\begin{figure*}[ht]
    \centering
    \includegraphics[width=\textwidth]{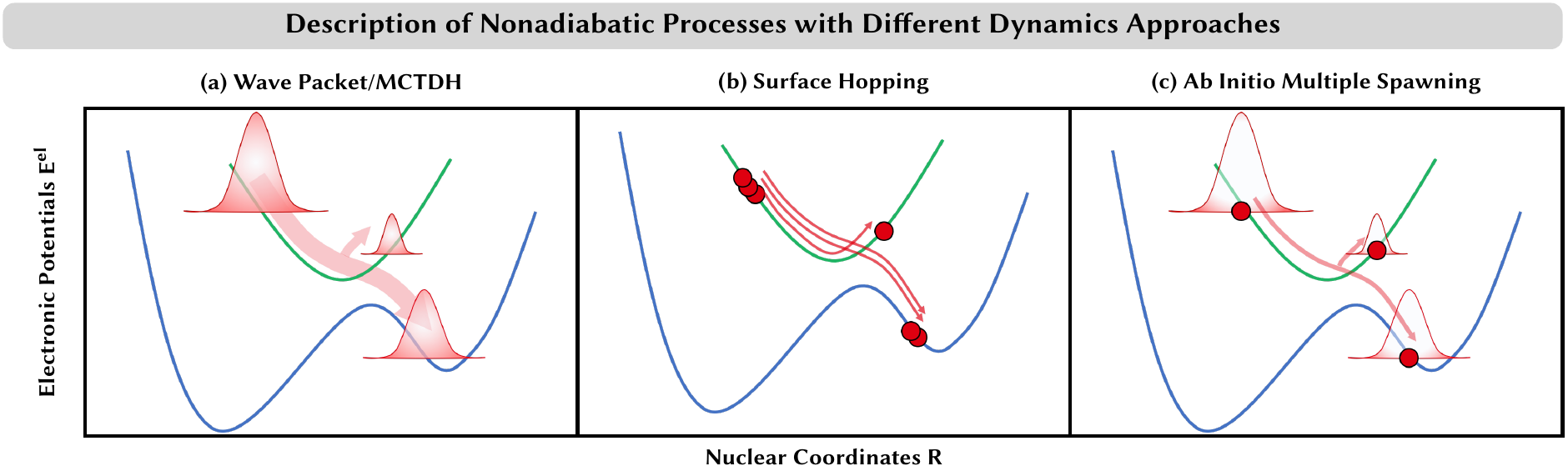}
    \caption{Schematic representation of selected approaches for nuclear nonadiabatic dynamics. (a) Wave Packet/MCTDH: a wave packet can split when electronic states come close in energy. The resulting wave packets remain coupled on different potential energy surfaces at all simulation times. (b) Surface hopping: several classical trajectories  (circles) are propagated independently. When the electronic states come close in energy, some trajectories hop stochastically to a different state, while the remainder stays on the same electronic state. (c) Ab initio multiple spawning: a Gaussian wave function follows a classical trajectory (depicted by the circle). When electronic states come close in energy, additional Gaussian wave functions can be spawned that remain coupled and can exchange amplitude as long as they are close together (small difference in nuclear coordinate $R$).}
    \label{fig:method}
\end{figure*}

This wave packet contains all quantum effects and can split in the presence of couplings.
It is usually discretized on a spatial grid along the $3N-6$ degrees of freedom of the PES of the molecular system, see Figure~\ref{fig:method}a. 
Assuming M grid points for each degree of freedom, a full dimensional wave packet propagation requires the pre-calculation of M$^{3N-6}$ grids points, which becomes quickly prohibitively expensive for all but the smallest molecules.
This \textit{curse of dimensionality} enforces the practical application of wave packet grid-based methods to very small systems or --in most of the cases-- to a selection of only few relevant nuclear degrees of freedom.

In the case of transition metal complexes, this meant performing wave packet dynamics simulations along one\cite{Finger1996JPC, Daniel1996IJQC,Daniel2001CP,Full2003PCCP,Daniel2003SCI,Full2006CP, Costa2008NJC,Ando2012CPL,Gourlaouen2015JCTC} or two dimensions.\cite{Daniel1994JPC,Daniel1994IJQC,Heitz1997JACS, Heitz1997CCR, Heitz1997JCP, Guillaumont1998CCR,Guillaumont1999JACS,  Daniel1993JPC, Heitz2000JOC,BruandCote2002CEJ, Ambrosek2007JCP, Ambrosek2007JPCA} 
Most of these studies\cite{Daniel1993JPC,Full2003PCCP,Daniel1994JPC, Daniel1994IJQC,Daniel1996IJQC, Finger1996JPC,Full2006CP,Heitz1997JACS, Heitz1997CCR,Heitz1997JCP,Guillaumont1998CCR,Guillaumont1999JACS,Heitz2000JOC,Daniel2001CP,BruandCote2002CEJ,Daniel2003SCI, Ambrosek2007JCP, Ambrosek2007JPCA} were concerned with ultrafast ligand dissociation in hydride and carbonyl complexes upon photoexcitation, for which the natural coordinate of choice was the bond distance between the metal and one or two of the ligands to detach. 
How to come up with an optimal low-dimensional coordinate space in general, for a molecule and reaction to study, is not necessarily an easy problem and different approaches are employed, see below Section~\ref{sec:wp_select}.
In passing we note that while some of this studies incorporated explicitly laser pulses to initiate the dynamics or even to guide it,\cite{Daniel2001CP,Daniel2003SCI,Ambrosek2007JCP,Full2006CP, Full2003PCCP} in others the laser-matter interaction is excluded and excitation is assumed to take place instantaneously (see Section~\ref{sec:prac_excite}).

\subsection{Multi-Configurational Time-Dependent Hartree}\label{sec:mcdth}
The multi-configurational time-dependent Hartree method (MCTDH)\cite{Beck2000PR} is another form of quantum dynamics, where the nuclear wave function is expanded in a set of single-particle functions (SPFs) $\varphi$ as 
\begin{equation}
    \Psi^{nuc}(R_1, R_2, \ldots, R_f, t)=\sum_{j_1=1}^{n_1}\ldots\sum_{j_f=1}^{n_f}A_{j_1\ldots j_f}(t)\prod_{\kappa=1}^f\varphi_{j_\kappa}^{(\kappa)}(R_\kappa, t)
    \label{eq:mctdh}
\end{equation}
where $A_{j_1\ldots j_f}(t)$ are the MCTDH expansion coefficients.
This approach is analogous to the MC treatment that was introduced in the electronic structure theory (see Section~\ref{sec:elec_mr}) with respect to a single Slater determinant.
One subtle difference between both fields is that for the nuclear problem, (symmetric) Hartree products are used in the configurations while the electronic problem requires (anti-symmetrized) Slater determinants, that change sign upon exchanging two particles, following the Pauli principle.
Similar to the case of propagating wave packets on a grid (Section~\ref{sec:wp_standard}), the computational cost of MCTDH also scales exponentially with the degrees of freedom included.
The important difference, however, is that in this case the exponential scaling is given by the number of SPFs per coordinate 
instead of the number of grid points. 
As the number of SPFs needed is smaller than the number of grid points, MCTDH is computationally more economic and therefore allows to consider a larger number of degrees of freedom than by propagating on a grid. 
In addition, a number of strategies have been devised in the last years to increase the number of dimensions, e.g. by combining several individual modes through using multi-mode SPFs \cite{Worth1998JCP} or in the multi-layer MCTDH variant.\cite{Wang2003JCP,Vendrell2011JCP}

\subsection{Vibronic Coupling Models}\label{sec:wp_lvc}

\begin{figure*}[ht]
    \centering
    \includegraphics[width=\textwidth]{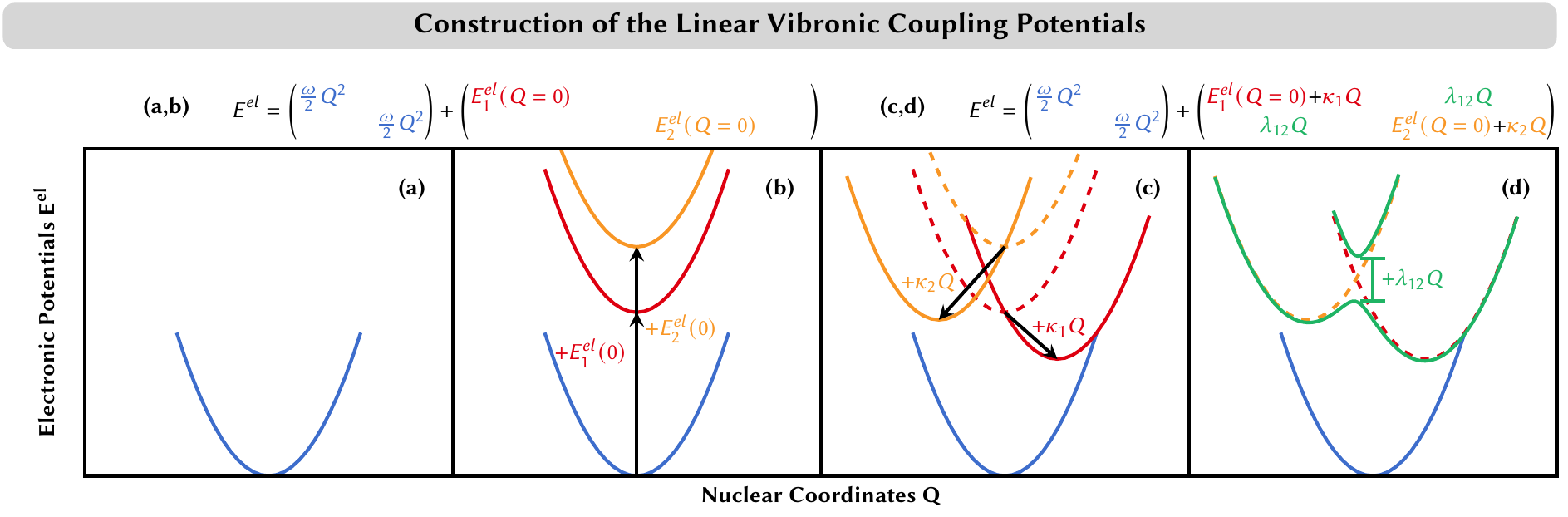}
    \caption{Construction of the linear vibronic coupling (LVC) potential. (a) Harmonic ground-state potential $E^{el}_0$ (blue curve). (b) Adding vertical excitation energies $E^{el}_i$ shifts harmonic excited-state potentials vertically (red, orange curves). (c). Adding intrastate coupling constants $\kappa_i$ shifts harmonic excited-state curves diagonally (from dashed to solid curves). (d) Adding interstate coupling constants $\lambda_{ij}$ couples the excited-state potentials (from dashed orange/red curves to solid green curves).}
    \label{fig:lvc}
\end{figure*}

An extended strategy to harness the efficiency of MCTDH is to employ vibronic coupling models\cite{Koeppel1984ACP} to describe the PESs on which the wave packets can be propagated.
In a vibronic coupling model, the PES are expanded in a Taylor series around a reference geometry $R=Q_0$ (usually the Franck-Condon geometry) using mass-frequency scaled normal coordinates $Q$, see Figure~\ref{fig:lvc}.
The Taylor series is often truncated after the first (linear) term --what is then known as the linear vibronic coupling (LVC) model --so that the PES $E^{el}$ are approximated in terms of the ground-state PES $E^{el}_0$ and linear vibronic coupling terms $W$
\begin{equation}
    E^{El}(Q)=E^{el}_0(Q)+W(Q)\label{eq:lvc_pot}
\end{equation}
The ground-state PES are approximated as harmonic oscillators with frequencies $\omega_i$
\begin{equation}
    E^{el}_0(Q)=\sum_{Normal\;modes\;i=1}^{3N-6}\frac{\hbar\omega_i}{2}Q_i^2 \label{eq:lvc_v0}
\end{equation}
The coupling terms read
\begin{equation}
    W_{nm}(Q)=\left\lbrace\begin{array}{cc}
         E^{el}_n(Q_0) + \sum_{i=1}^{3N-6}\kappa_i^{(n)}Q_i& \text{for}\quad n=m  \\
         \sum_{i=1}^{3N-6}\lambda^{(n,m)}_iQ_i& \text{for}\quad n\neq m
    \end{array}\right.\label{eq:lvc_coupling}
\end{equation}
where $E^{el}_n(Q_0)$ are the vertical excitation energies at the Franck-Condon geometry, while $\kappa_i^{(n)}$ and $\lambda_i^{(n,m)}$ are the intrastate and interstate couplings elements for the normal mode coordinate $Q_i$.

By definition, the usage of LVC models is limited due to the harmonic approximation of the potentials.
This approximation neglects anharmonic effects that can be essential in different situations, e.g., to describe torsional motion or dissociation.
Furthermore, due to parameterization, the LVC potentials can only describe nuclear motion in the vicinity of the reference geometry. 
Thus, LVC models work best in rigid molecules.

\begin{figure*}
    \centering
    \includegraphics[width=\textwidth]{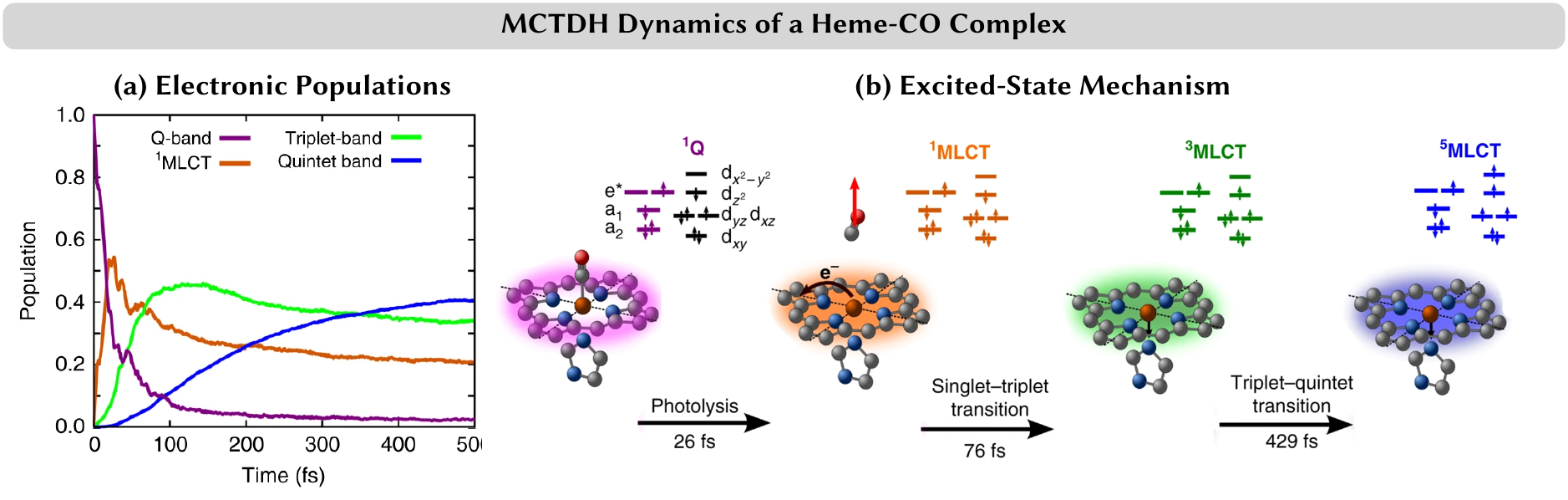}
    \caption{(a) Time evolution of the electronic state populations and (b) mechanism obtained from multi-configurational time-dependent Hartree (MCTDH) dynamics of a heme-CO complex.\cite{Falahati2018NATC}
    MCTDH simulations conducted on 15-dimensional potential energy surfaces based on CASSCF/CASPT2 linear vibronic coupling (LVC) models. 
    After initial excitation in the Q-band, a metal-to-ligand charge transfer state is populated that displays large-amplitude Fe-CO oscillations. 
    The system then further decays to the triplet and quintet manifolds.
    Adapted with from Ref.~\citenum{Falahati2018NATC} under Creative Commons Attribution 4.0 International License (visit  http://creativecommons.org/licenses/by/4.0/). Copyright by the Authors 2018. 
    }
    \label{fig:applications_mctdh}
\end{figure*}

Despite these limitations, LVC has become the standard approach to calculate PES in wave-packet dynamics simulations using MCTDH for transition metal complexes\cite{Worth2006MP, Falahati2018NATC, Fumanal2017JCTC, Fumanal2018JPCL, Fumanal2018PCCP, Eng2015ACR, Harabuchi2016JCTC, Papai2016JPCL, Capano2014JPCA, Papai2016JPCC, Papai2018JCTC, Papai2019JCP, Papai2019JPCC, Eng2020PCCP, Giret2021JMCC, Fumanal2021JCP} making it possible to include up to 16 nuclear degrees of freedom.\cite{Fumanal2021JCP}
One example of MCTDH using 15 degrees of freedom in a heme-CO complex is shown in Figure~\ref{fig:applications_mctdh}.
Among these studies, it is worth to mention that only two include explicitly the excitation by a laser pulse.~\cite{Papai2018JCTC, Papai2019JCP}

The convenience of LVC models is that the computational effort in the electronic structure step is mostly reduced to determining the coupling parameters and those can be obtained from a small number of calculations for each normal mode individually.

\subsection{The Choice of Degrees of Freedom}\label{sec:wp_select}

Due to the curse of dimensionality, regardless whether one performs wave packet propagation on a grid or using MCTDH as well as combined with LVC models, a crucial decision is always the selection of how many and which degrees of freedom need be considered, i.e., which are the most important coordinates that describe the problem at hand.
Going beyond natural dissociation coordinates and involving the degrees of freedom that connect the Franck-Condon region with one (or more) conical intersections and excited-state intermediates is not an easy task that in most cases goes beyond chemical intuition.

One approach thereby is to select normal modes based on the size of their vibronic coupling terms.\cite{Capano2014CHI, Capano2014JPCA, Eng2015ACR, Fumanal2021JCP} 
As large vibronic coupling elements are needed to efficiently transfer population between the electronic states, these coupling modes are necessary to describe the excited-state dynamics.
This selection can be extended by adding tuning modes, which are normal modes that are responsible for the largest displacements in the excited-state dynamics by reaching towards the excited-state minima\cite{Papai2016JPCL, Papai2016JPCC, Papai2018JCTC, Papai2019JCP, Papai2019JPCC} and the excited-state crossing points.\cite{Harabuchi2016JCTC, Fumanal2017JCTC, Fumanal2018PCCP, Fumanal2018JPCL} 

An additional strategy to identify an optimal coordinate subspace is to use more low-cost dynamics methods that allow including all or a very large amount of degrees of freedom and then identify a posteriori the most important ones. 
These can be then selectively considered in more accurate quantum dynamical approaches. 
To this category of low-cost methods belong ab initio molecular dynamics approaches, that will be introduced next. 

\section{Ab Initio Molecular Dynamics}\label{sec:aimd}

One alternative to wave packet-based dynamics is ab initio molecular dynamics (AIMD).
In AIMD, the nuclei are described as classical particles that follow  Newton's (classical) laws of motion on electronic potentials $E^{el}$ obtained by quantum-chemical methods
\begin{equation}
    M_A\frac{\partial^2}{\partial t^2}R_A(t) = -\nabla_{R_A}E^{el}(R(t))
    \label{eq:newton}
\end{equation}
This is only an approximate description of the nuclear motion.
In reality, the motion naturally follows the laws of quantum mechanics.
Accordingly, by definition, AIMD excludes nuclear quantum effects such as tunneling or  coherence in the nuclar motion, and can --at best --be corrected a posteriori. 
Describing the motion of the nuclei classically in AIMD, however, introduces a huge practical advantage for dynamics. 
As the motion of the nuclei follows a (classical) trajectory $R_A(t)$ that is at each time step determined only be the current molecular geometry, the (exponentially scaling) pre-computing part of the entire PES is lifted off.
Instead, the electronic-structure calculations can be performed "on-the-fly" during the dynamics simulation, whereby the necessary properties to propagate the nuclear dynamics such as electronic potentials and their gradients are only calculated at the current geometry.

We note that, in principle, it is also possible to generate a PES within certain approximations where to run classical trajectories.
In one case, a semiglobal PES of \ce{[Cu(phen)_2]^+} in solution was obtained from molecular dynamics trajectories, albeit on uncoupled S$_0$ and S$_1$ states.\cite{Agena2017CPL}
As it will be describe later, LVC models can also be used to run AIMD trajectories. 
In all cases though, the classical nature of the nuclei implies that a swarm of trajectories to be propagated is needed, instead of the one single propagation needed in wave packet dynamics. 

When several coupled electronic states are considered, two problems appear due to the nature of the classical trajectory approximation.
One is that, unlike the wave packet which can split in the presence of couplings (recall  Figure~\ref{fig:method}a), in AIMD a recipe is needed to transfer classical particles between different electronic states.
The other is that again, unlike a wave packet that spreads over different electronic states and each portion follows the gradient in its corresponding PES, in AIMD every classical particle is confined to a single point of the PES and follows a single gradient that has be to be decided somehow.

In the following two of the AIMD methods, which have been used up to now for excited state dynamics of transition metal complexes will be described. 

\subsection{Surface Hopping}\label{sec:sh}

Probably the most popular AIMD approach that includes a mechanism to transfer population between different electronic states is surface hopping (SH).\cite{Tully1990JCP, Tully1998FD}
In SH dynamics, the electronic wave function is allowed to spread over different electronic states as it is expressed as a linear combination of several electronic states,
\begin{equation}
    \Phi^{el}(r, t; R)=\sum_{Elec.\;States\;j}c_j(t)\phi_{j}(r;R)\label{eq:elec_lc}
\end{equation} 
Its time evolution is determined by the time-dependent electronic Schr\"odinger equation (in analogy to eq.~\ref{eq:tdse_nuc2})
\begin{equation}
    i\hbar\frac{\partial\Phi^{el}(r, t; R)}{\partial t} = \op{H}^{el}\Phi^{el}(r, t; R)\label{eq:tdse_el}
\end{equation}
which yields the time dependence of the coefficients $c_j(t)$
\begin{equation}
    \frac{\partial c_j(t)}{\partial t}=\sum_{Elec.\;States\;k}-\left( \frac{i}{\hbar} \mel{\phi_j}{\op{H}^{el}}{\phi_k}  +\mel{\phi_j}{\frac{\partial}{\partial t}}{\phi_k}\right)c_k(t)\label{eq:eom_elec}
\end{equation}
Note that in this standard formulation, the Hamiltonian already excludes the light-matter interaction and is restricted to the electronic states of the system.
The first term in the parenthesis in eq.~(\ref{eq:eom_elec}) is the coupling between the different electronic states, while the second term can be computed using the NAC between and electronic states and the velocity of the nuclei $v_R$
\begin{equation}
    \mel{\phi_j}{\frac{\partial}{\partial t}}{\phi_k} = \mel{\phi_j}{\nabla_R}{\phi_k}v_R\label{eq:nac_velo}
\end{equation}

The trajectory in SH follows the gradient of a single electronic state, the so-called active state, $\phi_i$.
After every time step in the simulation, the trajectory is allowed to change the active state, i.e., to "hop" to a different electronic PES (see Figure~\ref{fig:method}c) with a certain probability.
This probability is often calculated using the fewest-switches criterion,\cite{Tully1990JCP} which ensures a minimum number of hops along the propagation.
This criterion prevents the system from effectively travelling along an averaged gradient in the unfortunate case of a system hopping every time step.
The probability $P_{i\to j}$ for a hop from initial state $\phi_i$ to final state $\phi_j$ can be expressed as 
\begin{equation}
    P_{i\to j}=\frac{2\Delta t}{c^\ast_i(t)c_i(t)}\Re\left\lbrace c_i^\ast(t)c_j(t)\left[ \frac{i}{\hbar}\mel{\phi_i}{\op{H}^{el}}{\phi_j}+ \mel{\phi_i}{\frac{\partial}{\partial t}}{\phi_j} \right] \right\rbrace
    \label{eq:sh_prob}
\end{equation}
This equation shows that the probability for a hop can only become large in  the presence of large NACs --through the second term in brackets (cf. eq.~\ref{eq:nac_velo}) --and when the electronic wave function of the system has already sizable admixture of the final state --through the coefficient $c_j(t)$ and the thus necessary $c_i^\ast(t)c_i(t)<1$.

After a trajectory hops from one electronic state to another, its potential energy changes instantly. 
To conserve the total energy, its kinetic energy needs to be adjusted.
This is done by re-scaling the momenta of the nuclei, which, in practice, is best achieved by re-scaling along the direction of the NACs.\cite{Plasser2019JCTC, Barbatti2021JCTC}
A problem appears when a hop should occur according to the probabilities calculated from the electronic wave functions (eq.~\ref{eq:sh_prob}), but the trajectory has insufficient nuclear kinetic energy to compensate for the potential energy change during the hop.
In a fully quantum description, such a transition can be allowed due to the tunneling effect. 
However, in the classical description used in SH, such transitions --referred to as \textit{frustrated hops} --are not allowed.
Consequently, standard SH is not able to describe processes involving tunneling effects, although using a modified hopping criterion, it is possible to explore tunneling pathways qualitatively (see Section~\ref{sec:long_rare}).

In order to mimic a wave packet, AIMD methods employ a swarm of trajectories, which in SH are independent of each other.
While following a single path along the nuclear coordinates, the propagation of the electronic wavefunction still faces the problem that it is completely coherent.\cite{Schwartz1996JCP, Bittner1995JCP}
This means that all parts of the electronic wave function, even if they are in different electronic states, are all propagated along the same gradient, which is the gradient of the active state.
This description is erroneous. 
Instead, each part of a wave packet should experience the gradient of the electronic state that it occupies and be moved with individual velocities according to corresponding gradient, thus, losing the coherence of motion among them over time.  
As a remedy, SH simulations employ different types of so-called decoherence corrections.\cite{Subotnik2016ARPC, Plasser2019JCTC}
For example, in the easiest from all, the energy-based decoherence correction,\cite{Zhu2004JCP} the electronic populations on the non-active states are continuously damped at each time step.
The decoherence time that determines the rate of this damping thereby depends on the energy difference between the active and non-active states as well as the kinetic energy of the trajectory.\cite{Granucci2007JCP}
Thus, the larger the energy gap between the states and the faster the system moves, the faster the electronic populations decohere.

\subsubsection{The Cost of Surface Hopping Simulations}\label{sec:sh_cost}
The computational cost of a SH simulation is basically determined by the underlying on-the-fly electronic structure calculations. 
The total cost depends on the number of trajectories propagated.
However, as in SH all trajectories are independent, their calculation can be well parallelized.
A SH trajectory uses a typical nuclear time step of 0.5~fs. 
This means, for a total simulation time of, say, 500~fs, we need 1000 time steps and if a swarm of 100 trajectories is considered, this adds up to $10^5$ electronic structure calculations that need to be performed.
For reference, this number is comparable to the number of calculations necessary to pre-compute a five-dimensional PES with 10 grid points in each dimension where to propagate a wave packet. 
The computational advantage of SH methods explains why it has been extensively used in the last decades\cite{Wang2016JPCL,CrespoOtero2018CR} to study the excited states in a broad variety of organic materials and, to a lesser extent, also in transition metal complexes.\cite{Paterson2002JPCA,Tavernelli2011CP, Freitag2014IC,Atkins2017JPCL, Liu2018JPCA, Mai2019CS, Fang2019CTC, Talotta2020CEJ,Liu2018JCPb}
On passing we note that, from these studies, only one\cite{Mai2019CS} considered the effect of a laser excitation explicitly.

\begin{figure*}[ht]
    \centering
    \includegraphics[width=\textwidth]{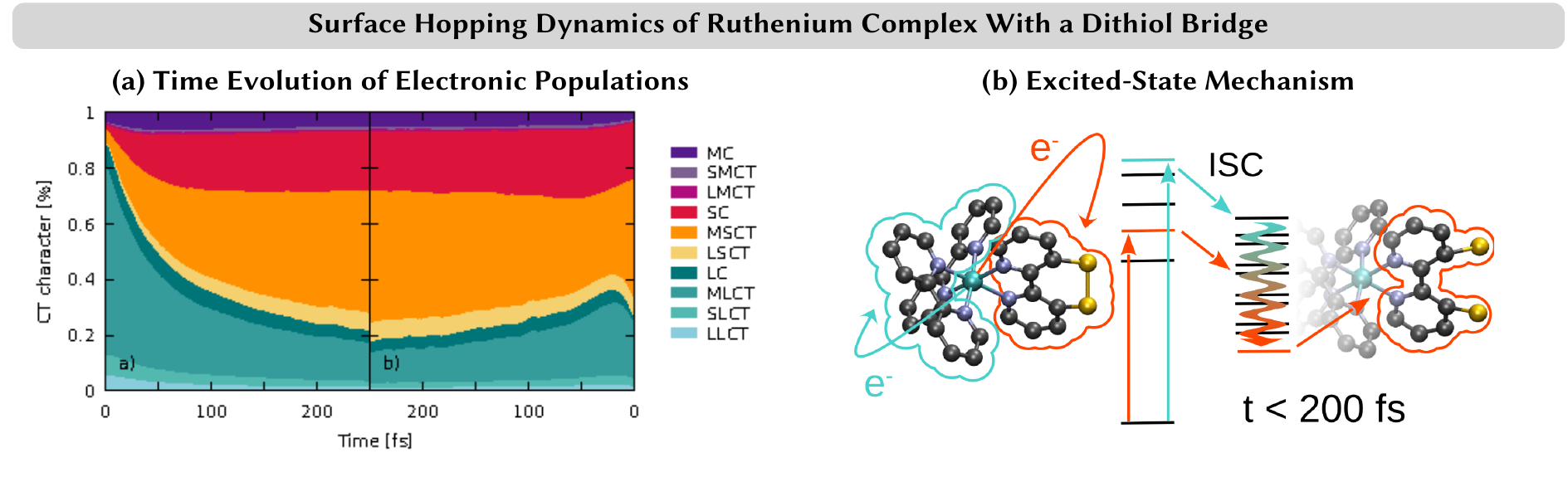}
    \caption{Excited state  dynamics of \ce{[Ru^{(II)}(bpy)$_2$(^{S-S}bpy)]^{2+}} (bpy  = 2,2'-bipyridine, \ce{^{S-S}bpy} = [1,2]Dithiino[4,3-b:5,6-b']dipyridine).\cite{Heindl2021IC}
(a) Time evolution of electronic state populations colored according to the charge transfer (CT) character. (b) Charge transfer mechanism obtained from a surface hopping study carried out on 161-dimensional potential energy surfaces based on TD-DFT linear vibronic coupling models. An excitation at a high-energy absorption bands that is dominated by metal-to-ligand charge transfer (MLCT) excitations to the bpy ligands [green/turquoise contributions in (a), starting at the left-hand-side] is after 250 fs very similar to the results obtained when exciting into the lowest-energy absorption band corresponding to states where the \ce{^{S-S}bpy} ligand [yellow/orange contributions in (a), starting at the right-hand side] is predominant. This means that in less than 200 fs excitations are located in the di-sulfide ligand, regardless of the excitation wavelength. Adapted from Ref.~\citenum{Heindl2021IC} under a Creative Commons Attribution 4.0 International license (visit: http://creativecommons.org/licenses/by/4.0/). Copyright by the American Chemical Society 2021.}
    \label{fig:applications_sh}
\end{figure*}

The implementation of the LVC model within SH\cite{Plasser2019PCCP} has enabled further efficiency and therefore a cheaper application to transition metal complexes.\cite{Dorn2020JACS, Zobel2020IC, Heindl2021IC, Mai2020TCA, Zobel2021CS}
In this way, it is possible to deal with systems with more than hundred nuclear degrees of freedom, propagate for several picoseconds, and consider thousands of trajectories --a previously inaccessible venture for on-the-fly SH.
Figure~\ref{fig:applications_sh} exemplifies the capabilities of SH dynamics using LVC potentials defined for 166 normal modes of Ru$^{(II)}$(bpy)$_2$($^{S-S}$bpy)]$^{2+}$ and using almost 9000 trajectories.\cite{Heindl2021IC} 
On the dark side, however, even with LVC, it is sometimes not possible to include all normal modes.
Especially low-frequency modes can experience non-physically large displacements when describing the motion of a molecule in the basis of normal modes, and these modes need to be excluded.\cite{Dorn2020JACS, Zobel2020IC, Heindl2021IC, Mai2020TCA}

\subsubsection{Exploiting Surface Hopping To Find Relevant Degrees of Freedom}\label{sec:sh_dof}

As discussed in Section~\ref{sec:wp_select}, the selection of nuclear degrees of freedom to be included in wave packet/MCTDH simulations can be challenging. 
An interesting approach is then to use a combination of wave-packet and SH methods.\cite{Capano2017PCCP,Gomez2019JPCA}
For instance, SH simulations were performed for a copper(I) complex in solution including all vibrational degrees of freedom.\cite{Capano2017PCCP}
Using a principal component analysis, the dominant normal modes activated during the SH excited state decay were identified, and could be used in a subsequent, more accurate wave packet dynamics simulation.

One way to identify the important normal modes that can obtained from a single simulation run is to follow the activity of each normal mode during the dynamics.
However, there exist also more sophisticated approaches that take into account the coupling of the nuclear motion to the evolution of the electronic state population.
For example, normal mode coherence or correlation analyses include the comparison of the motion of AIMD trajectories in excited states and in the ground state or monitor the effect of normal modes on excitation energies, energy gaps, and the overlaps between electronic state wave functions.\cite{Mai2019JCP}
Furthermore, the FrozeNM algorithm can be used to freeze normal modes and observe the effect that their exclusion has on the time evolution of the electronic states.\cite{NegrinYuvero2020JCTC}
Finally, a machine-learning algorithm has been developed that can identify global reaction coordinates in excited-state reactions from AIMD simulations in an automatic manner, given that the AIMD simulations provide sufficiently large data sets for statistical evaluation.\cite{Tavadze2018JACS}

As an alternative avenue, SH simulations performed using LVC potentials are so efficient that they can be gradually repeated reducing its dimensionality until the differences to the full dimensional calculation are acceptably small.
In this way, a minimum set of degrees of freedom can be identified, as illustrated in a \ce{[PtBr_6]^{2–}} complex that could be reduced from its 15-dimensional space initially considering 200 electronic states to a 9-dimensional problem with 76 electronic states without loss of accuracy.\cite{Gomez2019JPCA}

\subsection{Ab Initio Multiple Spawning}\label{sec:aims}

\begin{figure*}
    \centering
    \includegraphics[width=\textwidth]{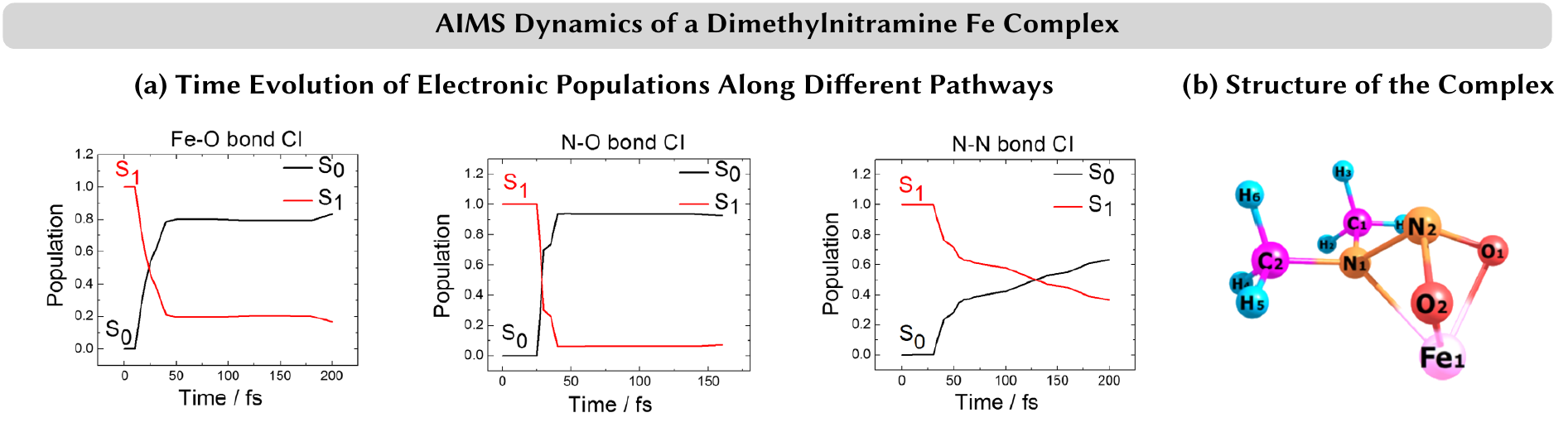}
    \caption{(a) Time evolution of electronic state populations of different pathways of ab initio multiple spawning (AIMS) dynamics of a dimethylnitramine iron complex (b). 
    AIMS simulations conducted on 33-dimensional potential energy surfaces calculated on-the-fly using CASSCF. 
    Three different reaction channels corresponding to \ce{Fe-O}, \ce{N-O}, and \ce{N-N} bond dissociation could be identified.
    Adapted with permission from Ref.~\citenum{Bera2017JCP}. Copyright by the Authors 2017.    }
    \label{fig:applications_aims}
\end{figure*}

A number of other methods to study excited state dynamics exist, which in terms of cost and approximations  can be placed formally between the MCTDH and SH formalisms.\cite{Gonzalez2021QCES}
From them, only the ab initio multiple spawning (AIMS) approach\cite{BenNun2000JPCA} has been employed in the excited state dynamics of transition metal complexes, an Fe-dimethylnitramine complex,\cite{Bera2017JCP} see Figure~\ref{fig:applications_aims}.

AIMS is derived from full multiple spawning (FMS)\cite{Martinez1996JPC}, in which Gaussian functions --also referred to as trajectory basis functions (TBF) --are propagated on classical trajectories.
When TBFs enter regions with high probability to transfer population between different electronic states (regions with strong NACs), new TBF are spawned on the electronic states that are encountered, and both the initial as well as the spawned TBF are propagated further, see Figure~\ref{fig:method}b.
In contrast to other methods using classical trajectories, FMS requires the pre-computation of the complete PESs to mediate the coupling between the TBF, and it is formally exact.

In AIMS, the couplings between the TBFs are calculated only locally around the regions where the TBF are getting close to each other. 
Using this approximation, AIMS simulation can also be performed on-the-fly. 
Furthermore, AIMS simulations usually employ the independent-first-generation approximation, in which all initial (parent) TBF are independent; only the spawned (child) TBF stay coupled to the initial TBF.
This approximation can be justified by assuming that the nuclear wave packet will usually spread rapidly in phase space in the beginning of the dynamics, which then would allow to neglect the coupling between the parent TBFs.
This approximation is exaggerated further in SH --substituting spawns by hops --in which there is no coupling at all between the trajectories.
It could likewise be justified by assuming that the nuclear wave packet spreads rapidly also after non-adiabatic events.
However, in practice, this approximation has proven to be flawed and resulted in the introduction of decoherence corrections in SH.

With the parent and child TBFs coupled in AIMS, the computational demand scales quadratically with the number of TBFs, thus, making AIMS more costly than SH when comparable numbers of trajectories and TBFs are used.
The higher scaling may be alleviated by introducing a stochastic-selection scheme, in which spawned TBFs can be removed.\cite{Curchod2020JPCA} 
When the coupling between TBFs becomes small, one of the coupled TBFs is selected, the population of the other TBFs is collapsed into the selected TBF, and only the selected TBF is propagated further.
In this manner, approximate AIMS simulations could be run at similar cost as SH simulations with results close to that of standard AIMS\cite{Ibele2021JCP} --so far without spin-orbit couplings\cite{Curchod2016JCP} and  not yet applied to any transition metal complex. 
Furthermore, AIMS simulations can be made to describe tunneling dynamics by allowing to spawn TBF in the same electronic state, e.g., when the distance between the tunneling particle and its donor particle surpasses a certain threshold.\cite{BenNun2000JCP}

\section{Environmental Effects}\label{sec:gap_environment}
\begin{figure*}
    \centering
    \includegraphics[width=\textwidth]{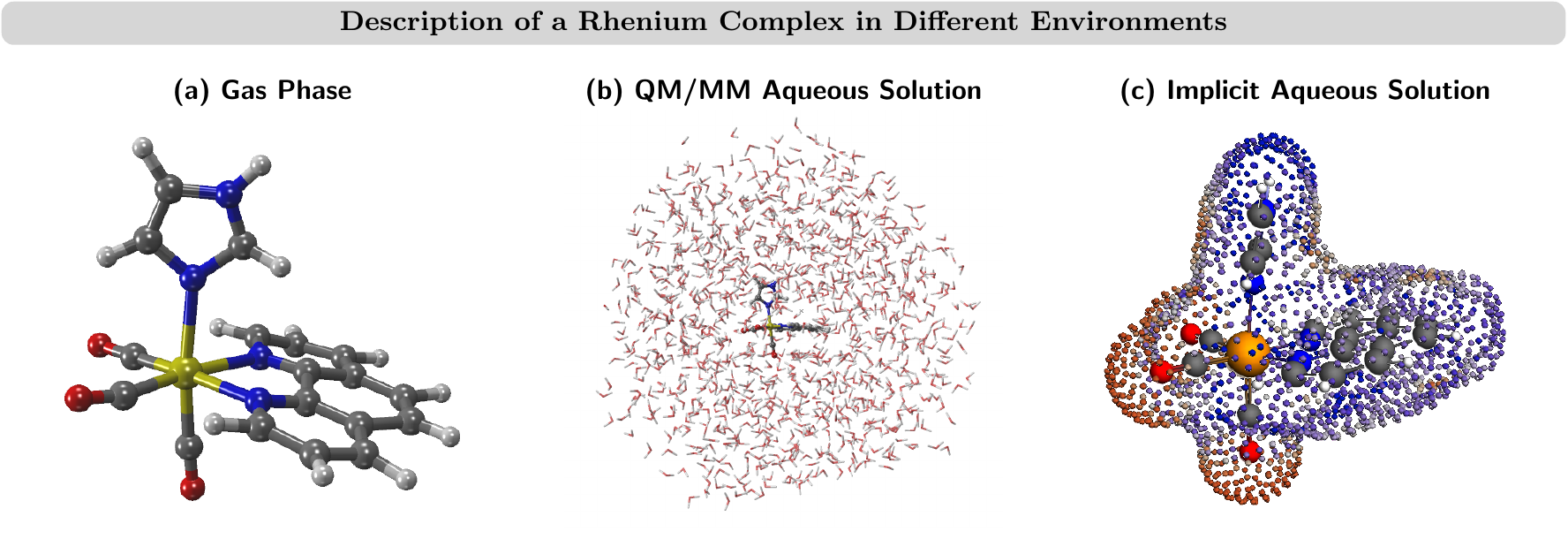}
    \caption{Description of the \ce{[Re(CO)_3(phen)(im)]^+} complex in different environments (phen = phenanthroline, im = imidazole). (a) In gas phase. (b) In QM/MM aqueous solution surrounded by 1054 water molecules. (c) In implicit aqueous solution. Little spheres represent point charges on the cavity surrounding the complex. Figure adapted from Ref.~\citenum{Mai2017PCCP} under a Creative Commons Attribution 3.0 Unported Licence. Published by the PCCP Owner Societies (2017).}
    \label{fig:solvent}
    \end{figure*}

Very often, the phenomena one is interested in occurs in an environment, being either a solvent or a biological surrounding structure. 
Accordingly, dynamics simulations of transition metal complexes should be simulated in the same media. 
In the following, we discuss two approaches that are readily used for excited state dynamics: in one the environment is included explicitly and in the other, typically a solvent is only accounted for implicitly, see Figure~\ref{fig:solvent}.

\subsection{Explicit Environments: QM/MM Partitioning}\label{sec:gap_qmmm}

In Section~\ref{sec:elec_semi}, we discussed the introduction of semi-empirical approximations in the electronic structure calculations in order to make calculations of large molecules feasible. 
When a large amount of, for example, solvent molecules should be included in the calculation, semi-empirical methods are rarely enough. 
In these cases, the system can be further approximated with a partition in two (or more) regions, where one of them --at least the transition metal complex-- is treated quantum mechanically (QM) and the rest is only accounted for with molecular mechanics (MM), i.e., replaced by parameterized force fields.\cite{Adcock2006CR}
Force fields contain classical energy expressions for bond lengths, bond angles, and dihedral angles as well as long-range interaction terms such as van-der-Waals and electrostatic interactions.
This classical approach is computationally very economic, and it allows to simulate dynamics of systems with $>10^6$ atoms.
Therefore, combined with the QM methods to describe the electronic excited states, it is ideal to treat transition metal complexes in solution or in an biological environment.

Depending how the interaction between the regions is defined, different QM/MM methods exist.\cite{Senn2009ACIE, Brunk2015CR}
In all schemes, the computational effort is mostly due to the expense of the QM part of the calculation.
Therefore, the size limits for a typical QM region in hybrid QM/MM calculations is not larger than the molecular size limit for the QM computation of the isolated molecules.
Due to the structural flexibility of biological environments or solvents, PES cannot be characterized by unique points such as global energy minima or minimum-energy crossing points in QM/MM calculations. 
Instead, several thermally accessible minima that can be populated exist and should be properly sampled,\cite{Nogueira2018ARPC} increasing the complexity of the calculations.

QM/MM approaches have been applied in nonadiabatic AIMD simulations of several transition metal complexes in solution.\cite{Capano2017PCCP, Tavernelli2011CP, Mai2019CS, Fang2019CTC}
In all these studies, the transition metal complex is described alone in the QM region, while the bulk of the solvent ($\sim10^3$ molecules) are described using MM force fields, as in Figure~\ref{fig:solvent}b. 
Although QM/MM-AIMD simulations of organic systems embedded in biological environments exist,\cite{Li2010JPCL, Groenhof2008JACS, Groenhof2004JACS, Fingerhut2012JCP, Weingart2011PCCP, Polli2010NAT} we are not aware of any example with a transition metal complex. 

Including environment effects using QM/MM approaches in wave packet dynamics is also possible, however, it requires more elaborate schemes.
These schemes can include the modification of pre-computed PES of the isolated molecule with energetic shifts calculated from ground-state QM/MM-MD simulations\cite{Thallmair2015JCTC, Reiter2018JACS} or an iterative update of the solvent effects that is obtained from the simultaneous simulation of the solvent in an ensemble of classical trajectories.\cite{Cerezo2018JCTC}

\subsection{Implicit Environments}\label{sec:gap_implicit}
An alternative approach to incorporate environmental effects 
is to replace their atomistic description 
by a dielectric continuum.
This approach was created and is particularly useful to account for solvation.\cite{Mennucci2012WIRES, Mennucci2013PCCP}
Popular implementations of this approach include the polarizable continuum model (PCM)\cite{Miertus1981CP} and related variants, such as conductor-like screening model (COSMO)\cite{Klamt1993JCSPT2} or the SMD model.\cite{Marenich2009JPCB}
In these models, the solute molecule is placed inside a cavity containing charges on its surface, through which the interaction between the solute molecule and the surrounding solvent continuum is described (see Figure~\ref{fig:solvent}c). 

These models have been applied in excited-state dynamics of solvated transition metal complexes parameterized with LVC potentials using both, MCTDH\cite{Eng2015ACR, Harabuchi2016JCTC, Eng2020PCCP, Gourlaouen2015JCTC, Fumanal2017JCTC, Fumanal2018PCCP, Fumanal2018JPCL, Papai2019JPCC, Fumanal2021JCP} and AIMD simulations.\cite{Zobel2020IC, Dorn2020JACS, Heindl2021IC, Mai2020TCA}
Although, in principle, there exist stationary studies where also biological environments are approximately modelled with continuum models with very small dielectric constants, such strategy has not been used for dynamics, as it cannot capture the explicit fluctuations of the complex environments. 

Something to keep in mind when describing an excitation process using a continuum model, is that the solvent effects can be split into two contributions: a fast, dynamical component and a slow, inertial component.\cite{Mennucci2012WIRES}
The fast component describes the interaction of the electron density of the solvent with the electron density of the solute in its excited state, that can be considered changed instantaneously after excitation.
The slow component takes into account the situation that the solvent molecules --though modelled as a dielectric continuum --are still oriented referring to the initial electron density of the ground state.
When describing the dynamical evolution after photoexcitation, the solvent effects will still be well approximated by the slow component referring to the ground state, where they consist mainly on the modulation of the energy differences between the different electronic states.\cite{Santoro2021PCCP}
However, at later times in the dynamics, when the solvent molecules start to adopt their orientation to the changed electron density in the excited state, this approximation becomes worse.

\section{About Initial Conditions, Zero-Point Energy, and Temperature}\label{sec:gap_init}

Excited-state dynamics simulations of nuclear motion requires the selection of an initial condition that describes the state of the nuclei 
before the excitation process. 
In wave packet/MCTDH dynamics, the initial wave function is typically the lowest vibrational state of the electronic ground state.   
When the PES are expressed in an LVC model, the initial wave function can be described by the ground-state wavefunction of an harmonic oscillator.
Starting the system in the vibrational ground state $\Psi_0$ corresponds to describing the system at zero temperature, and the total energy of the system is then given by its zero-point energy (ZPE).
Such situation is adequate to mimic gas phase experiments performed in a cold beam, as for example those in described in Ref.~\citenum{Paterson2002JPCA,Daniel2003SCI,Full2003PCCP, Full2006CP}

When the system adopts a finite temperature, vibrationally excited states 
become populated with temperature-dependent probabilities
, that can be obtained, e.g., from statistical sampling.\cite{Lorenz2014JCP} 
Despite its simplicity, there are no wave packet/MCTDH dynamics simulations of transition metal complexes including temperature.

In AIMD simulations, the classical nuclei are described by their position and momenta, which need to be selected as initial conditions.
Two common approaches exist to this purpose: Wigner sampling (also referred to as quantum sampling) and molecular dynamics sampling (classical sampling).\cite{Barbatti2016IJQC, Zobel2018JCTC}
In the Wigner sampling, coordinates and momenta are sampled from a Wigner distribution\cite{Wigner1932PR} --a simultaneous probability distribution of coordinates and momenta, that is a function of the nuclear wave function.
For this, one typically employs the harmonic-oscillator approximation for which analytical expressions of the nuclear wave functions are known.\cite{Sun2010JCP}
Wigner sampling is commonly employed in the zero-temperature formalism, where the total energy of the system is again given by the ZPE.
However, the effects of finite temperature can easily be accounted for by allowing the population of excited vibrational states.\cite{Zobel2019PCCP} 
In the classical sampling, initial position and momenta are taken from snapshots from MD trajectories propagated in the electronic ground state.
In the ground-state MD simulation, the system is given a total energy of $k_BT$ and the simulations are typically performed for at room temperature ($T=300$~K).
This energy --$k_BT$ at room temperature --is much smaller than the ZPE. 
The inclusion of the different amounts of vibrational energy from both sampling approaches can have an effect, e.g., at reaction rates in excited state dynamics.\cite{Barbatti2016IJQC}
Since the total energy at room temperature is better approximated by the ZPE, (zero-temperature) Wigner sampling is, thus, usually preferable over MD-based sampling.
When studying large systems in a hybrid QM/MM scheme --which cannot be described completely by QM --it is also possible to combine Wigner sampling for the QM part and classical sampling for the MM part.\cite{Ruckenbauer2010JPCA,Mai2018FC}

\section{Including Explicit Light-Molecule Interactions}\label{sec:prac_excite}

\begin{figure*}[ht]
    \centering
    \includegraphics[trim=7.00cm 0 0.75cm 0, clip, width=8.5cm]{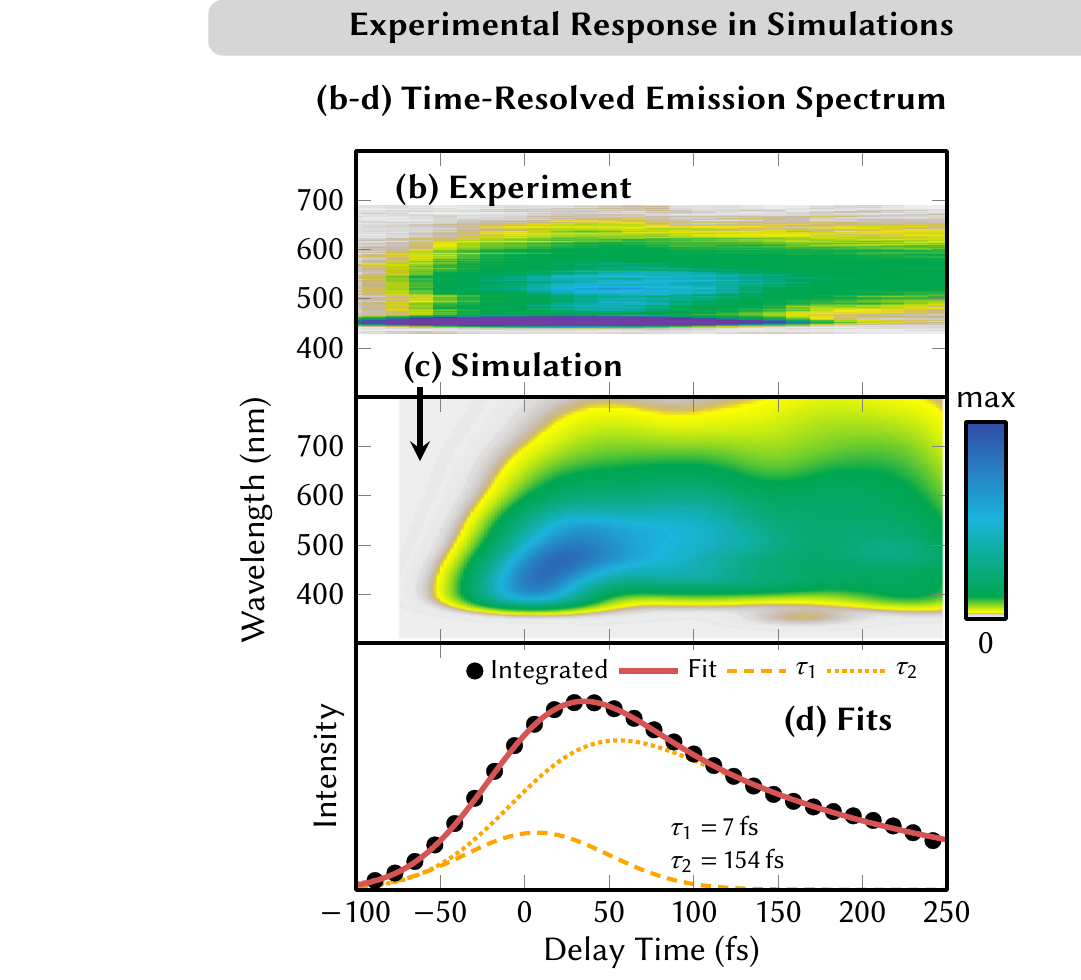}
    \caption{ (b-d) Comparison of experimental and simulated time-resolved emission spectra of \ce{[Re(CO)_3(im)(phen)]^+} in aqueous solution. im = imidazole; phen = phenanthroline. (b) Experimental time-resolved luminescence signal\cite{ElNahhas2011IC} for \ce{[Re(CO)_3(im)(phen)]^+} (c) Same observable computed from surface hopping trajectories.\cite{Mai2019CS} (d) Integrated simulated spectrum (black dots) and biexponential fit. The obtained time constant of 154~fs is in excellent agreement with that obtained experimentally (144~fs). Adapted from Ref.~\citenum{Mai2019CS} under a Creative Commons Attribution 3.0 Unported Licence. Published by The Royal Society of Chemistry 2019.}
    \label{fig:exp}
\end{figure*}

One goal of nonadiabatic dynamics simulations is to study the processes occurring after excitation of the molecule into its excited electronic states.
In spectroscopic experiments, this excitation is usually realized by irradiation of the molecule with a laser pulse centered around a specific wave length, that is usually known as the initial \textit{pump} pulse.
Furthermore, time-resolved experiments need a second delayed pulse to \textit{probe} the excited state reaction in time. 
This pulse sequence is then known as a pump-probe experiment. 
Probe pulses can ionize the molecule or excite as well as de-excite it to other electronic excited states.

Naturally, it would be desirable that the simulations imitate the experiment as close as possible, in particular when a comparison or interpretation of a particular experiment is aimed at. 
Formally, the laser pulses can be easily added to the Hamiltonian (cf. Eq.~\ref{eq:hamil}) and in the field of quantum dynamics, there exist a number of studies of transition metal complexes using explicit laser pulses.\cite{Daniel2001CP, Full2003PCCP,Ambrosek2007JCP,Daniel2003SCI, Papai2018JCTC, Papai2019JCP}

As time-resolved spectroscopic experiments typically use weak laser pulses that only excite a few percent of the total population, the simulation of the excitation process can be cumbersome, particularly in AIMD simulations. 
In wave packet dynamics propagations, one would choose laser parameters similar to the experimental ones and make sure that the corresponding small populations do not run into numerical problems.
In AIMD, although laser interactions can be simulated,\cite{Richter2011JCTC,Mitric2011PCCP,Mignolet2016JCP} the use of laser pulses that only excite a tiny fraction of the molecules is impractical\cite{Persico2014TCA} because most of the calculated initial conditions are disregarded.  
For that reason, explicit laser excitations to excite initially the molecule are ignored in AIMD simulations of transition metal complexes --with exceptions.\cite{Mai2019CS}
In general, for simplicity in wave packet dynamics and for economic reasons in AIMD simulations, the excitation is only considered implicitly through the appropriate selection of the initially excited electronic states.
We note, however, that the simulation of the time-resolved experimental signal can be computed \textit{a posteriori} from AIMD simulations, using the available trajectories to excite or ionize the system further. 
Figure~\ref{fig:exp}b-d shows one example of a simulated time-resolved luminescence spectra signal for a rhenium complex in solution obtained with SH simulations.

The excitation process in wave packet dynamics is often simulated through a vertical projection of the nuclear wave packet from the electronic ground state into the potential of the absorbing electronic state. 
This approach is valid in the limit of ultrashort laser pulses (so-called $\delta$-pulses), where the nuclear wave packet in the excited state is simply a copy of the vibrational ground state in the electronic ground state scaled by the transition dipole moment.\cite{Schinke1993}
At the other extreme, the simulation of long laser pulses (a continuous wave) corresponds to exciting the molecule into a vibrational state of its excited electronic state.
In reality, a laser pulse with a finite duration will achieve an intermediate situation, that is best taken into account if explicitly included in the simulation. 
We note that care has to be taken when including the laser pulse explicitly in the case of exciting into a set of degenerate electronic states, as this can lead to dynamics that are dependent on the polarization of the laser in the simulation.\cite{Papai2018JCTC}

In AIMD simulations, the ultrashort-pulse limit for the initial excitation is usually emulated by computing the excited states of a large set of nuclear geometries (sampled as described in Section~\ref{sec:gap_init}) and then selecting the states and associated geometries by the size of the oscillator strength.\cite{Barbatti2007JPPA, Persico2014TCA}
In order to keep a reasonable number of initial conditions, for practical purposes, the initially excited states are selected not at a specific wave length but within an energy window around the intended excitation energy. 
As AIMD simulations require a large number of trajectories for statistical averaging, accordingly, a much larger number of trial states need to be generated to start trajectories only in states with (relatively) large oscillator strengths.
Typical excitation energy windows range from 0.2~eV\cite{Heindl2021IC, Dorn2020JACS} to 0.5~eV\cite{Zobel2020IC, Talotta2020CEJ, Atkins2017JPCL, Zobel2021CS}.
In some cases, this selection scheme is also approximated by starting trajectories in one specific state based on its large oscillator strength at the Franck-Condon geometry.\cite{Liu2018JPCA, Fang2019CTC, Tavernelli2011CP, Freitag2014IC, Bera2017JCP}

It is also possible to think about an excitation process different from the coherent light coming from the laser pulses typically employed in femtosecond spectroscopy: the interaction of molecules with natural thermal light coming from sources such as the sun --as encountered in biological processes.\cite{Brumer2018JPCL}
Besides the broader spectral excitation range, natural thermal light is stationary incoherent irradiation and may, thus, trigger other responses than observed when using coherent pulsed laser irradiation.\cite{Brumer2012PNAS}
The incoherent nature of natural thermal light during the excitation process can be described through an ensemble of coherent pulses, whereby all effects of incoherence are recast in the post-excitation averaging.\cite{Chenu2016JCP}
The effects of stationary incoherent irradiation can then be implemented in nonadiabatic dynamics simulations through running an ensemble of dynamics simulations, displacing the individual runs in time and using a weighted averaging scheme over all runs.\cite{Barbatti2020JCTC}

\section{Bridging Time Scales}\label{sec:long}
\subsection{State of the Art}\label{sec:long_now}

\begin{figure*}
    \centering
    \includegraphics[width=\textwidth]{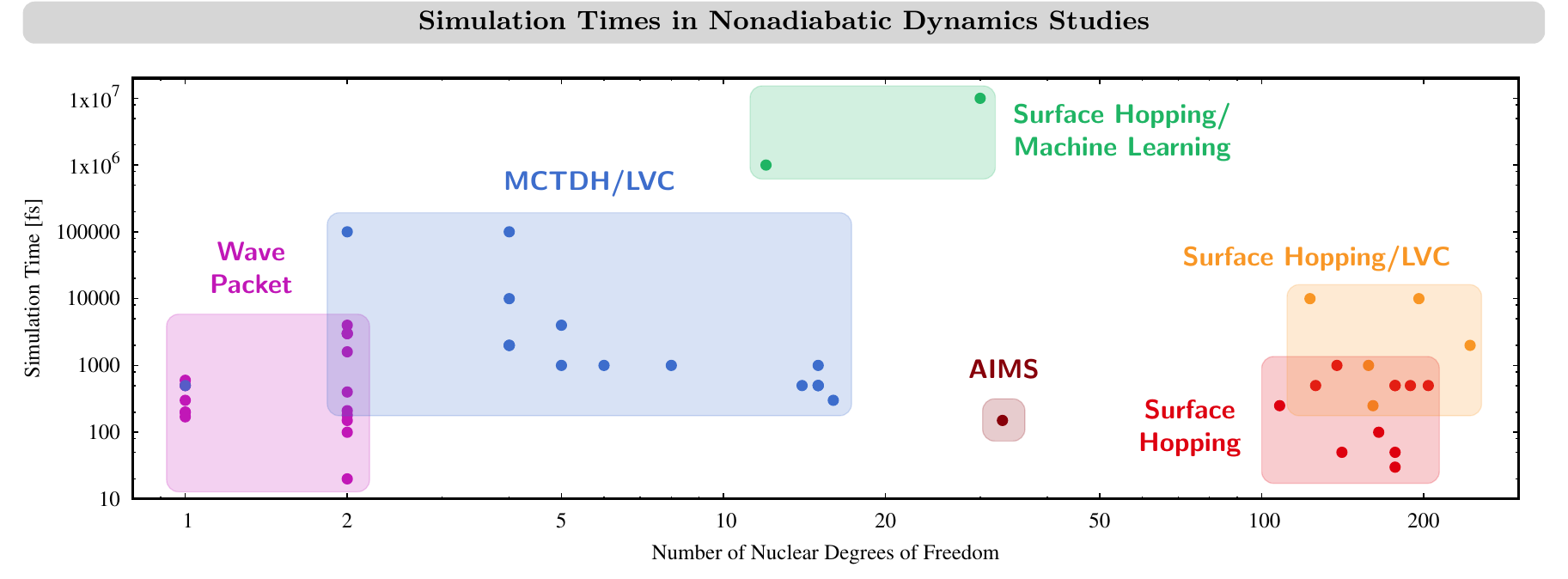}
    \caption{Simulation times (colored dots) in nonadiabatic dynamics using different methods as a function of the number of nuclear degrees of freedom. Note the double-logarithmic scale. Shaded areas denote typical combinations based on the literature surveyed in this Perspective and do not refer to any strict limits. All simulation times taken from studies of transition metal complexes, except surface hopping/machine learning studies which were applied to organic molecules.\cite{Westermayr2019CS, Li2021CS} MCTDH = multi-configurational time-dependent Hartree. AIMS = ab initio multiple spawning. LVC = linear vibronic coupling. }
    \label{fig:times}
\end{figure*}

With the standard techniques described until now, the time scales that can effectively be simulated using nonadiabatic dynamics are basically determined by their computational cost (see an overview in Figure~\ref{fig:times}).
For wave packet propagations (recall Section~\ref{sec:wp}), this cost is hidden in the pre-calculation or parameterization of the PES and scales exponentially with the number of nuclear degrees of freedom.
Accordingly, nowadays, simulation times for transition metal complexes range from 1~ps on 15-dimensional PES\cite{Fumanal2017JCTC} over 4~ps on 5-dimensional PES\cite{Papai2019JCP} to 100~ps on two-dimensional PES.\cite{Eng2020PCCP}
In on-the-fly AIMD methods (Section~\ref{sec:aimd}), the computational cost is dominated by the electronic structure calculation that is performed at every time step.
A typical simulation time is 1~ps, e.g., in a ruthenium(II) complex with 138 nuclear degrees of freedom including three singlet and three triplet states.\cite{Talotta2020CEJ}

If using parameterized potentials, the time span of AIMD simulations can be substantially extended.
For example, using LVC parameterized potentials, SH were run for several picoseconds for iron(II) and vanadium(III) complexes on 244 and 197-dimensional PES including total numbers of 60 and 25 electronic states, respectively.\cite{Zobel2020IC, Dorn2020JACS} 
In general, the cost of SH simulations on LVC potentials is low enough to perform even longer dynamics simulations. 
However, since the parameterization of the LVC potentials is usually conducted at the Franck-Condon geometry, the description of the dynamics becomes worse the farther the system moves away from the Franck-Condon geometry. 
Thus, simulations based on LVC potentials become less reliable for longer simulation times -- both using AIMD and wave packet dynamics.
In the following we describe several scenarios that envision how to bridge the gap of large time scales and what can expected to be soon used for the excited state dynamics of transition metal complexes.

\subsection{Machine Learning}\label{sec:long_ml}
A promising alternative to extend the time scales is to exploit parameterized PES from 
machine learning (ML) algorithms,\cite{Dral2020JPCL,Westermayr2020MLST, Westermayr2021CR} although this approach has been only applied to relatively small organic molecules.
ML algorithms use parameterized functions that have been fitted to a reference test set --the training set --in order to predict properties of interest for new data points that are not in the training set.
For example, ML algorithms were trained on excitation energies and oscillator strengths at a set of different ground-state geometries and used to predict these properties at other ground-state geometries and thus simulate the complete absorption spectrum.\cite{Ye2019PNAS, Xue2020JPCA}

The first applications of ML algorithms to nonadiabatic dynamics simulations have been reported on on-the-fly computed PES.\cite{Richings2017JCTC,Hu2018JPCL, Dral2018JPCL, Chen2018JPCL, Westermayr2019CS, Li2021CS, Ha2021JCTC}
Studies thereby highlighted difficulties to describe the dynamics in regions of strong NACs or close to conical intersections when using ML with kernel ridge regression algorithms.\cite{Hu2018JPCL, Dral2018JPCL}
In contrast, dynamics based on ML featuring neural networks seem to handle these problematic regions better.\cite{Ha2021JCTC, Chen2018JPCL, Westermayr2019CS, Li2021CS}

The ability of the ML algorithm to predict properties depending on the training data introduces two problems in the simulation of nonadiabatic dynamics.
First, in order to describe molecular properties correctly in the all regions of the PES that the molecule visits during the dynamics, all regions must be taken into account in the training of ML algorithm.
For large, polyatomic molecules, the knowledge about which regions are important is usually not available a priori. 
This problem can, however, be solved within the ML algorithm itself through adaptive sampling.\cite{Westermayr2019CS, Li2021CS}
In this approach, two or more ML algorithms are trained independently and their predictions are compared during the simulation. 
When the differences between their predictions at a certain molecular geometry surpass a predefined threshold, the geometry is assumed to lie in a region of insufficient training data.
The properties at this geometry are then calculated using a reference method, the data is added into the training data, and the ML algorithm can, thus, be improved iteratively.
In this way, dynamics on parameterized dynamics can be simulated also for long-time scales up to nanosecond in a meaningful manner.\cite{Westermayr2019CS, Li2021CS}

The second problem concerns the choice of the reference method for the training data.
As the ML algorithm is trained to reproduce the results of the reference method, the choice of the reference method defines the ultimate accuracy of the ML algorithm.
Since typical training sets need thousands of training points, the compilation of the training data can itself become computationally expensive.
However, also this task can be alleviated using ML techniques, i.e., $\Delta$ machine learning.\cite{Ramakrishnan2015JCTC, Ramakrishnan2015JCP}
In $\Delta$-ML, an algorithm can be trained to predict corrections to computationally inexpensive electronic structure methods through comparison to more accurate reference data.
This approach, however, has not yet been applied in combination with excited state dynamics simulations.

\subsection{Metadynamics}\label{sec:long_speeding}
Metadynamics is a technique invented to accelerate rare events occurring in the electronic ground state.\cite{Laio2002PNAS} 
In meta-surface hopping\cite{Lingerfelt2016JCTC} it is possible to speed-up the dynamics and sample faster reaction pathways by accelerating the rate 
of nonadiabatic transitions.
Since this rate depends linearly on the nuclear kinetic energy 
and quadratically on the NACs, 
it can be accelerated by scaling up the NACs; this has been first demonstrated in the relaxation rate from the $S_1$ state of a small organic molecule.\cite{Lingerfelt2016JCTC}

A related approach to meta-surface hopping is given by multi-state metadynamics,\cite{Lindner2018PRA, Lindner2019JCTC} which extends the original (ground-state) metadynamics approach\cite{Laio2002PNAS} to multiple electronic states.
In metadynamics, the dynamics of a molecule are started in its ground state equilibrium, and successively external potentials are added to drive the molecular dynamics towards a desired region.
For multi-state metadynamics, this desired region is the conical intersection between the ground and the first excited state.
Thus, instead of monitoring the decay to the ground state starting from the excited state, the possible reaction pathways are searched in reverse direction.

\subsection{Rare Event Sampling}\label{sec:long_rare}
The rate of a process in nonadiabatic dynamics depends on the probability that it occurs. 
In wave-packet dynamics simulations, low-probability processes occurring on a long timescale can be captured by extending the simulation time (within reason), as the wave packet at every time step is allowed to spread over all accessible reaction pathways.
In AIMD simulations, such as SH, where trajectories follow a single pathway, describing such rare events can be difficult, as besides the cost of the long simulation, a sufficiently large ensemble of trajectories is required. 
The margin of error of a random statistical sampling is approximately given by
\begin{equation}
    \varepsilon_\gamma=z_\gamma(N_{Trajectories})^{1/2}\label{eq:moe}
\end{equation}
where $z_\gamma$ is the quantile of the standard deviation.
This means that e.g. for a 95~\% confidence interval ($z_{0.95}=0.98$), a process that accounts for 10 or 1~\% of the reaction yield requires ca.~100 or $10^4$ trajectories, respectively, for statistical meaningful results.
In order to sample even rarer events, the army-ants algorithm has been developed for SH\cite{Nangia2004JCP}, where trajectories are stochastically driven to explore also low-probability channels.
This significantly reduces the number of trajectories needed to converge results for simulating processes with very low probabilities ($\sim10^{-8}$).
The army-ants algorithm has also been used to model tunneling paths employing classical trajectories.
This is done by pushing the trajectories to go beyond the classical turning points in ground-state molecular dynamics simulations\cite{Zheng2014CS} and mean-field Ehrenfest dynamics including excited states.\cite{Zheng2014JPCL, Xu2014JACS}

\subsection{Rate Methods}\label{sec:long_rate}
One example of a process that can be slow is intersystem crossing. 
Fast intersystem crossing events can be monitored by doing excited state simulations including SOCs (cf. Eq.~\ref{eq:hij}).
However, when intersystem crossing is too slow, time scales can be obtained indirectly from static calculation of intersystem crossing rates\cite{Moitra2021PCCP, Penfold2018CR, Marian2012WIRES}, e.g., based on Fermi's golden rule
\begin{equation}
    k_{ISC}=\frac{2\pi}{\hbar}\sum_{Final\;States\; f}\left|\mel{\Psi^{mol}_i}{\op{H}^{SO}}{\Psi^{mol}_f} \right|^2\delta(E_i-E_f)\label{eq:fermi}
\end{equation}
The molecular wave functions $\Psi^{mol}$ are thereby taken as a product of an electronic and a nuclear part, the latter typically approximated by harmonic oscillators, and all coupling elements are calculated at a reference geometry for which usually the equilibrium geometry of the initial state $\Psi^{mol}_i$ is selected.
This approach has been used to calculate intersystem crossing rates for a number of transition metal complexes.\cite{Luedtke2020PCCP, Foeller2016IC, Kleinschmidt2015JCP, Heil2016MP, Moitra2018PCCP, Paul2017JPCL, Sousa2013CEJ}
Due to the selection of a reference geometry where the intersystem crossing is to occur and the assumption that the system resides in an initial state at the beginning of the intersystem crossing process, these static approaches are limited in the processes that they are able to describe.
Thus, most of the applications\cite{Luedtke2020PCCP, Foeller2016IC, Kleinschmidt2015JCP, Heil2016MP} focus on the intersystem crossing connecting the lowest-excited singlet and triplet states $S_1$ and $T_1$, while few others\cite{Paul2017JPCL, Moitra2018PCCP} also take into account the $T_2$ state.
A different strategy was adopted in Ref.\citenum{Sousa2013CEJ}, where intersystem crossing rates were calculated for a network of several singlet, triplet, and quintet states of an iron (II) complex. 
In this case, the calculations were performed at the ground-state equilibrium geometry, which was deemed reasonable based on the ultrafast nature of the intersystem crossing established from experimental studies, i.e., giving the system only few tens of picosecond to move away from the ground-state equilibrium geometry after excitation before ISC occurs.

\section{Conclusions and Challenges Ahead}\label{sec:conclusion}
The field of excited state simulations on transition metal complexes is now about three decades old.
It started with exact quantum dynamics simulations in reduced dimensionality (one or two degrees of freedom) but it has been only very recently that it could made the leap to include all the important degrees of freedom or even full dimensionality in selected complexes.
The latter was possible either by exploiting parameterized potentials or/and in combination with on-the-fly methods, such as surface hoping or multiple spawning. 
With these techniques established, one can expect a booming in the field that will help rationalizing new experiments and making exciting predictions comparable to the pace that organic photophysics and photochemistry experienced in the last thirty years. 

However, the limitations of the methods are still plenty and many are the challenges ahead.
Even today, the dynamical methods that can be applied to transition metal complexes could be basically grouped at two extremes: one that rely on quantum nuclei but sacrifices degrees of freedom and another that can include all the degrees of freedom but sacrifices the quantum nature of of the nuclei. 
Unfortunately, as of today, the best of both worlds is only possible for the smallest organic systems but not for transition metal complexes. 
These means that often, depending on the molecule or even its electronic structure (which a priori is difficult to know), one strategy or the other can be more suitable (or the only one viable) and this is not a light-hearted decision to take.

One additional aspect is the inclusion of explicit light-matter interactions in the calculations, be it incoherent light sources such as the sun, a coherent excitation with a laser, the simulation of a full pump-probe experiment or even the application of tailored (optimal) laser pulses aimed to control nuclear and electronic properties of a transition metal complex. 
While early studies based on wave packets could describe these interactions easily (albeit in reduced dimensionality), the methods based on trajectories that can to-date be applied to transition metal complexes still face important difficulties in this endeavour.\cite{Mignolet2019JPCA,Heindl2021JCP} 

On top, one should realize that the suitability of the underlying electronic structure method is always key for success --irregardless of the chosen dynamical strategy --and its choice can hamper the efficiency and feasibility of the dynamical method itself, as well as the quality of the final results. 
For instance, some transition metal complexes might require multi-reference methods, 
which, while they might be available for stationary calculations, are not affordable for on-the-fly dynamical methods.
Thus, if the applicability of the multi-reference method in reduced-dimensional models is also too expensive or too simplistic (despite its application also being an intrinsic challenge itself!), one is left with no choice for dynamics to look at.  

In this Perspective, we have reviewed every dynamical study of transition metal complexes (to the best of our knowledge) and we could see that the list of "must-haves" for the next years is very long.
We are awaiting to see recently developed electronic-structure methods, such as DMRG, combined with any method for nuclear dynamics, to see more sophisticated on-the-fly techniques extended to transition metal complexes, to see more systematic studies where spectroscopic observables are included and can be directly compared with experimental signals, to see the capabilities of machine learning potentials applied to transition metal complexes, and studies based on metadynamics or rare event sampling on this field. 
Of course, simulations should be also be extended to include the description of the environment, beyond solution, as there are a lot of interesting applications of transition metal complexes embedded DNA, proteins, and other biological environments, as well as part of molecular machines and other supramolecular assemblies at the nanoscale.
A considerable challenge is also posed by polynuclear coordination complexes, where two or more metals are simultaneously light-active and responsible for the photophysics of the system. 
Besides the growth of the electronic structure problem, such complexes display magnetic exchange interactions, whose description with quantum chemical methods pose a challenge by itself. 
For such complicated problems, the advent of quantum computation\cite{Cao2019CR} could bring a new perspective by exploiting their unique features of superposition and entanglement.
First steps in the use of quantum algorithms for non-adiabatic dynamics\cite{Ollitrault2020PRL} have already been proposed.
We thus believe that the next decade will see an increase of fundamental new ideas to harness the excited state dynamics of transition metal complexes for the benefit of society and humankind in many areas.

\section*{Conflicts of interest}
There are no conflicts to declare.

\section*{Acknowledgements}
The authors thank S.~Mai for fruitful discussions and the University of Vienna for continuous support.
This work was financed by the Deutsche Forschungsgemeinschaft [DFG, Priority Program SPP 2102 “Light-controlled reactivity of metal complexes” (GO 1059/8-1)]. 

\bibliography{./main}

\end{document}